\documentclass[prb,aps,12pt,nofootinbib]{revtex4}
\usepackage{graphicx}
\usepackage{subfigure}
\newcommand{\be}{\begin{eqnarray}}
\newcommand{\ee}{\end{eqnarray}}

\begin{document}

\title{Smoluchowski  Dynamics and the Ergodic-Nonergodic Transition}
\author{Gene F. Mazenko }
\affiliation{The James Franck Institute and the Department of Physics\\
The University of Chicago\\
Chicago, Illinois 60637, USA}
\maketitle


\centerline{Abstract}

We use the recently introduced  theory for the kinetics of systems
of classical particles to investigate systems driven by
Smoluchowski dynamics.
We investigate the existence of
ergodic-nonergodic (ENE) transitions near the liquid-glass transition.
We develop a self-consistent perturbation theory in terms of
an effective two-body potential and work to second order in this potential.
At second order, we have an explicit relationship between the static structure factor
and the effective potential and choose the static structure factor
in the case of hard spheres to be given by the solution of the
Percus-Yevick approximation for hard spheres.  Then, using the analytically
determined ENE equation for the ergodicity function, we find an
ENE transition for packing fraction $\eta$ greater than a critical
value $\eta^{*}=0.76$ which is physically unaccessible.
The existence of a linear fluctuation-dissipation theorem in the problem
is shown and used to great advantage.



\newpage

\section{Introduction}

We continue\cite{FTSPD} our presentation of a self-consistent approach to the
kinetics of classical systems of particles by studying the fluctuations
in equilibrium of a system driven by Smoluchowski dynamics.
We first show that the system obeys a linear
fluctuation-dissipation theorem (FDT).
This simplifies the structure
of the theory significantly giving the conventional
 linear relation between the
density-density correlation function and the conjugate response
function.
We present here a perturbation theory valid to second order
in an effective interaction
potential.  Because of the self-consistency we are able to show that this
expansion is useful even for systems with hard-core interactions.

This approach was demonstrated at first order in Ref. \onlinecite{FTSPD}
(henceforth referred to as \mbox{FTSPD})
where the effective potential is found to be proportional to the direct
correlation function.  Here
we extend the calculation to second order.  In this paper we focus on the
collective (or Ornstein-Zernike)
second order self-energies.  We show that these self-energies
are quadratic functionals of the full density-density
correlation functions, the components of the self-energies
individually satisfy a FDT,  no wave-number or frequency cut offs
are needed\cite{ABL}, and the set of Dyson equations fundamental to
the theory can be replaced by a single-kinetic equation
of the same form as that produced in memory-function theories
\cite{MEMFUN, Crisanti}.

Going further, one can show that the collective contribution, again
using the FDT, gives a relation for the static structure which
agrees with the result from a purely static calculation
giving the structure factor in terms of the direct correlation
function and the direct correlation function self-consistently
in terms of the static structure factor.  Our approach is to
assume the static structure factor is known and to solve for
that effective potential which satisfies the second order
structural equation.  This effective potential is then available
in the dynamical calculations.  It could also be used in other
static calculations. We focus here primarily on the large wavenumber
regime near the structure factor maximum.

We also show that it is only the collective part of the self-energy
which enters into the determination of the ergodic-nonergodic (ENE)
phase-diagram.  The  ENE transition occurs when
the density-density correlation develops a zero-frequency
$\delta$ function. The amplitude of this $\delta$ function satisfies
a nonergodicity equation.  This equation
is very similar to that found\cite{MCT1} in mode-coupling theory (MCT).
For a system characterized by a structure factor obtained as
the solution of the Percus-Yevick approximation\cite{PY} for hard-spheres
we find a transition at packing fraction $\eta^{*} =0.76$.
Conventional MCT gives $\eta =0.51$. As obtained here, $\eta^{*}$ is well above any
physically attainable density in agreement with experiment and simulation.
If we drop self-consistency and use the first order effective potential
in determining $\eta^{*}$ we obtain the value $\eta^{*}=0.60$ even though
the first and second order effective potentials are very similar.
Therefore $\eta^{*}$ is a rather sensitive quantity. Despite our prejudices
that there is no physical ENE transition in the single-component hard-sphere
Smoluchowski dynamics, more work checking perturbations to our solution is required before one can
claim a ``proof" of the lack of a transition.

Experience guides us to be skeptical\cite{CrisantiPC}
 about the stability of this
transition under the inclusion of higher order terms in perturbation theory.
Happily one can test this skepticism by going to higher
order in the expansion.  This appears to be technically feasible.

In  a companion paper we discuss the {\it single-particle}
contribution to the second order self energies.  This term in
the second order self-energy makes no contribution to
the structural statics (static structure factor) but does
involve the equation of state.  It also does not come into the
nonergodicity-equation determining the ENE phase diagram.  This does
not mean that it plays no role in the slow dynamics of the system.
Quite to the contrary, this term depends linearly on $G_{\rho\rho}$
and in a way which suggests the $F_{12}$ model of Goetze\cite{F12MODEL} and
a mechanism for stretching the dynamics.

We find that the self-energy $\Gamma_{BB}$ is a functional,
to arbitrary order in the effective potential, of the full-density-density
correlation function. Setting aside the difficult
question of vertex renormalization,
the construction of $\Gamma_{BB}$ to arbitrary order appears
feasible. This involves construction of the self-energy
$\Gamma_{BB}$ as a polynomial in $G_{\rho\rho}$ which produces
a zero-frequency $\delta$ function in response to the
a zero-frequency $\delta$ function  developing in $G_{\rho\rho}$
in the nonergodic state.

We are interested in the kinetics of colloidal suspensions.
They are of interest because one can apparently carry out
clean experiments  in the regime where the system becomes glassy
or crystalizes\cite{CRYSTAL}.
The system is also of interest because the colloidal
particles are buffeted by a bath of smaller
particles which has the effect of rapidly thermalizing the
momentum degrees of freedom. This process is carefully described
by Fokker-Planck (FP) dynamics.  In FP dynamics the Newtonian
equations of motion are
supplemented by noise terms.  It is very convenient to study
a simpler dynamics than Fokker-Planck or Newtonian dynamics\cite{ND}.
To a first approximation in colloids we can assume that the
momenta thermalize quicker than the position variables.
Smoluchowski dynamics\cite{SD} assumes that the momenta are thermalized
and particles move via a random walk or diffusion process
interacting with the same two-body forces as in the Newtonian
case.  In the noninteracting limit
one has $N$-Brownian \cite{Brownian}
particles.

While the Smoluchowski dynamics offer a complete
self-consistent dynamical system with a static structure equivalent
to that for Newtonian and FP dynamics, there has been a search for simpler,
coarse-grained field theories applicable to the colloidal system.
Considerable energy has been focused on the
Dean-Kawasaki (SDK) model\cite{DEAN,KAW1}. The MSR action for this model
\cite{MSR1} is given by
\be
A_{DK}=\int d1\left[ D\rho(\nabla\hat{\rho})^{2}
+i\hat{\rho}\left[\frac{\partial \rho}{\partial t}
-\nabla\bigg(\rho\nabla\frac{\delta F}{\delta \rho}\bigg)\right]\right]
\ee
where $\rho$ is a density continuous field and $\hat{\rho}$ is its
response conjugate.
The functional derivative of the  effective
free energy, taken from density functional theory, is given by
\be
\frac{\delta F}{\delta \rho (x)}= ~T \ln \rho (x)
+\int d^{d}y ~ u(x-y)\delta \rho (y)
\ee
where the ideal gas contribution is proportional to $\ln \rho$
and $u$ is related to the direct correlation function.
An apparently appealing feature of this model is that the ideal-gas
contribution to the equation of motion satisfies
\be
\nabla \bigg(\rho\nabla\frac{\delta F_{IDG}}{\delta \rho}\bigg)
=T\nabla^{2}\rho.
\ee
There has been considerable effort to study the SDK system\cite{ABL}.
The most complete
analysis  is due to
Kim and Kawasaki\cite{KKrev}.
Analysis of the SDK model field theory uncovers multiple complications such as nonpolynomial nonlinear structure and multiplicative noise structure. This leads to an impractical nonlinear form of FDT relating response
and correlation
functions. In the end, one is not definitively able to answer
the question: Is there an ENE transition at one-loop order for the
SDK model?

Our second order self-consistent microscopic theory
suffers none of these problems and gives a definitive answer
to the ENE transition question.  An interesting point is how the theory
handles convergence of integrals in perturbation theory.
The theory naturally organizes itself into a structure with
self-dressed propagators.  This leads to convergent
integrals in perturbation theory.
One finds that
self-dressed propagators occur order to order in perturbation
theory. The dressing consists in multiplication of the physical
correlation function by
factors of $G^{(0)}$ and the effective potential.
It is crucial that one recognize that
these self-dressed correlation functions themselves
satisfy a FDT.

It is extremely useful
 that the linear FDT holds in the theory.
This facilitates the use of a simple kinetic equation in treating
the time evolution and separating out the static behavior.

While not immediately apparent,
the understanding of the role of one-particle irreducible
vertex functions in
this analysis is very important.  The expansion parameter is not, as in
conventional field theory, in terms of the vertices.
There are nonzero nonlinear vertices even at zeroth order in
the interaction. An example is the cubic 1PI vertex.
The one-particle irreducible vertices
are interesting even in the noninteracting limit.
Starting at the four-point  vertices one must deal
with one-particle reducible contributions to the cumulant structure.
Vertices are now generally frequency dependent.  We reserve the
discussion of the four-point vertices to the next paper in
this series.

The static theory can be
developed in complete analogy with the dynamic theory.   In
particular one can work out the self-consistent static expansion
in terms of the effective potential.  One finds  exactly the same results
for the statics using dynamics and statics.

The approach taken here is fully microscopic and allows one to calculate
in detail.  This is in contrast with the projection operator technique which
allows one to develop a useful phenomenological description of a
problem.
There was a substantial effort\cite{VCCA} to show that
a noninteracting  Brownian gas can be described by a nontrivial
but solvable cubic MSR field theory.  Our work has been
compatible with these results finding agreement for
the set of Brownian-density cumulants.

\section{Smoluchowski Dynamics}

Let us begin by defining the dynamical
system of interest.  Consider a set of $N$ particles with configurations
specified by the coordinates $R_{i}$
which satisfy the equations of motion
\be
\dot{R}_{i}= DF_{i}+\eta_{i}
\label{eq:2}
\ee
where the particles experience force
\be
F_{i}=-\frac{\partial}{\partial R_{i}}U(R),
\ee
with total potential
\be
U(R)=\frac{1}{2} \sum_{i\neq j}\bar{V}(R_{i}-R_{j})
\label{eq:4}
\ee
where $\bar{V}$ is a pair potential.
There is a noise source $\eta_i$ for each coordinate
which is taken to be Gaussian  with variance
\be
\langle\eta_{i}(t)\eta_{j}(t')\rangle =2k_{B}T D\delta (t- t ')\delta_{ij}
\ee
where $D$ is a diffusion coefficient.

We showed in FTSPD that one can set up a useful self-consistent
perturbation theory if we develop the theory in terms
a small set of collective variables
$\Phi$.  For this system, the density $\rho$ is essential
since it governs the static equilibrium behavior and, from the point
of view developed here, is always included in $\Phi =(\rho ,\ldots)$.
The set $\Phi$ must also include a response field
$B$ (described below) if we are to control and manipulate the interactions
in the system.
The set of collective variables treated [$\Phi =(\rho , B, \ldots)$],
is flexible and controlled by pairing
each observable with a conjugate external field [$H =(H_{\rho}, H_B,\ldots)$]
Here we specifically treat fluctuations in equilibrium
and choose $[\Phi =(\rho , B)]$.
We assume the system
is in equilibrium initially [$t=t_{0}$, $R_{i}(t_{0})=R_{i}^{(0)}$]
and the initial distribution
for a set of $N$ particles is canonical:
\be
P_0[R^{0}]=e^{-\beta U (R^{(0)})}/Z_{0}
\ee
where $U$ is the
potential energy defined by Eq.(\ref{eq:4}) and
$\beta$ is the inverse temperature.
The MSR action for the problem is given by
\be
A_{R }=\int_{t_{0}}^{\infty}dt_{1}
\sum_{i=1}^{N}\left[\hat{R}_{i}(t_{1})k_{B}T D \hat{R}_{i}(t_{1})
+i\hat{R}_{i}(t_{1})\cdot\left (\dot{R}_{i}(t_{1})
-D F_{i}(t_{1})\right) \right]
+A_{J}
\label{eq:8}
\ee
where the
contribution to the action $A_{J}$ is from the notorious
Jacobian\cite{JAJZJ}.

It was shown in FTSPD that this action can be written in the
highly compact form
\be
A =A _{0} +A_{I}
\ee
where $A_{0}$ is the quadratic part of the action
excluding the quadratic contribution to the
initial probability distribution,
\be
A_{0 }=\int_{t_{0}}^{\infty}dt_{1}
\sum_{i=1}^{N}\left[\hat{R}_{i}\bar{D}  \hat{R}_{i}
+i\hat{R}_{i}\cdot\dot{R}_{i} \right],
\label{eq:19}
\ee
where $\bar{D}=k_{B}TD$
and where the interaction is given by
\be
A_{I}=\frac{1}{2}\sum_{\alpha,\nu}\int d1d2
\Phi _{\alpha}(1)\sigma _{\alpha\nu} (12)\Phi _{\nu}(2).
\ee
The Greek labels range over  $\rho$ and $B$ and we introduce
the interaction matrix
\be
\sigma_{\alpha \nu} (12)=-\beta V(12)
\left[ \hat{\rho}_{\alpha}\hat{\rho}_{\nu}\delta(t_1-t_0)
-\beta^{-1}\left(\hat{\rho}_{\alpha}\hat{B}_{\nu}
+\hat{B}_{\alpha}\hat{\rho}_{\nu},
\right)\right]
\label{eq:IM}
\ee
where we have introduced the useful notation
\be
\hat{\rho}_{\alpha}=\delta_{\alpha \rho},
 \hat{B}_{\alpha}=\delta_{\alpha B}
\ee
and
\be
\bar{V}(12)=\bar{V}(x_{1}-x_{2})\delta (t_{1}-t_{2}).
\ee
The conjugate field is given by
\be
B(1) =D\sum_{i=1}^{N}\left[(\hat{R}_{i}i\nabla_{1}
+\theta (0)\nabla_{1}^{2})\right]
\delta (x_{1}-R_{i}(t_{1})).
\ee

The canonical partition function can be written in the
convenient form
\be
Z_{N}=\int \prod_{i=1}^{N}\bigg[{\cal D}(R_{i}){\cal D}(\hat{R}_{i})
d^{d}R_{i}^{(0)}\bigg]P_0(R_{i}^{(0)})
e^{-A_{0}-A_{I}+H\cdot\Phi}
\ee
\be
=Tr^{(N)} e^{-A_{I}+H\cdot\Phi},
\label{eq:23}
\ee
where we have introduced the average
\be
Tr^{(N)}{\cal O} =
\int \prod_{i=1}^{N}\bigg[{\cal D}(R_{i}){\cal D}(\hat{R}_{i})
d^{d}R_{i}^{(0)}\bigg] P_0(R_{i}^{(0)}) e^{-A_{0}}{\cal O}(R).
\ee
In the grand canonical ensemble,
\be
Z_{T}[H]=\sum_{N=0}^{\infty}\frac{\rho_{0}^{N}}{N!}
Tr^{(N)} e^{\int d1 H(1)\cdot\Phi (1)}
e^{\frac{1}{2}\int d1 d2 \Phi(1)\cdot\sigma\cdot \Phi(2)},
\label{eq:31a}
\ee
with the generator of cumulants given by
\be
W[H]= \ln Z_{T}[H].
\ee

\section{Self-Consistent Development}

It was shown in FTSPD that the one-point average
\be
G_{i}=\langle \Phi_{i}\rangle
=\frac{\delta}{\delta H_{i}}W[H]
\ee
satisfies the identity
\be
G_{i}=\tilde{T}r \phi_{i} e^{H\cdot\phi +\Delta W [H]},
\label{eq:45a}
\ee
where $i$ labels space, time and
fields $\rho$ or $B$, and where
\be
\Delta W[H] =W[H+F]-W[H]
\ee
with
\be
F_{i}=\sum_{j}\sigma_{ij}\phi_{j}
\label{eq:45}
\ee
and
\be
\Phi_{i}=\sum_{\alpha =1}^{N}\phi_{i}^{(\alpha)}.
\ee
We have
\be
\phi_{\rho}^{(0)}(1)=\delta[(x_{1}-R^{(0)}(t_{1})]
\ee
and
\be
\phi_{B}^{(0)}(1)=D[\hat{R}^{(0)}(t_{1})i\nabla_{x_{1}}+
\theta(0)\nabla_{x_{1}}^{2}]\delta [x_{1}-R^{(0)}(t_{1})].
\ee
These results were established in FTSPD using functional
methods.

The
dependence of the theory  on the interaction
potential is controlled by the quantity
$\Delta W[H]=W[H+F]-W[H]$.  We can expose the dependence on the potential by
constructing the functional Taylor-series expansion
\be
\Delta W[H] =\sum_{i}F_{i}\frac{\delta}{\delta H_{i}}W[H]
+\sum_{ij}\frac{1}{2}F_{i}F_{j}\frac{\delta^{2}}{\delta H_{i}\delta H_{j}}W[H]+\cdots
\ee
and we can conveniently introduce the set of cumulants
\be
G_{ij\ldots k}=\frac{\delta}{\delta H_{i}}\frac{\delta}{\delta H_{j}}
\ldots \frac{\delta}{\delta H_{k}}
W[H]
\ee
to obtain
\be
\Delta W[H] =\sum_{i}F_{i}G_{i}
+\sum_{ij}\frac{1}{2}F_{i}F_{j}G_{ij}
+\sum_{ijk}\frac{1}{3!}F_{i}F_{j}F_{k}G_{ijk}
+\ldots
\ee
with $F_{i}$ given by Eq.(\ref{eq:45}).
Clearly, in this form we can take $\Delta W$ to be a functional of $G_{i}$.
One can then use functional differentiation to express higher
order cumulants in terms of the one- and two-point
correlation functions $G_{i}$ and $G_{ij}$.

Of particular interest is that we established in FTSPD
a dynamic generalization of the static Ornstein-Zernike
relation\cite{OZR}.  Taking the functional derivative of Eq.(\ref{eq:45a}),
we have
\be
G_{ij}=\frac{\delta}{\delta H_{j}}G_{i}
\nonumber
\ee
\be
={\cal G}_{ij}+\sum_{k}c_{ik}G_{kj}
\label{eq:47}
\ee
where
\be
{\cal G}_{ij}=\tilde{T}r \phi_{i}\phi_{j}e^{H\cdot\phi +\Delta W}
\label{eq:54}
\ee
is a single-particle quantity and we have the memory function\cite{TDDC},
self-energy,  or dynamic direct correlation function given by
\be
c_{ij}=\tilde{T}r \phi_{i}e^{H\cdot\phi +\Delta W}
\frac{\delta}{\delta G_{j}}\Delta W.
\label{eq:34}
\ee
Since $\Delta W$ can be treated as a functional of $G_{i}$ we see
at this stage that we have available a self-consistent theory.
If we define the matrix inverses
\be
\sum_{k}\Gamma_{ik} G_{kj}=\delta_{ij}
\label{eq:56}
\ee
and
\be
\sum_{k}\gamma_{ik}{\cal G}_{kj}=\delta_{ij}
\label{eq:39}
\ee
then the two-point vertex is given without approximation as
\be
\Gamma_{ij}=\gamma_{ij}+K_{ij},
\label{eq:50}
\ee
where
\be
K_{ij}=-\sum_{k}\gamma_{ik}c_{kj}
\label{eq:38}
\ee
is the collective contribution to the self-energy.

\section{Fluctuation-dissipation theorems}

\subsection{Introduction}

The theory developed in FTSPD is very general and applicable
to a wide variety of nonequilibrium problems.
Here we look at fluctuations in equilibrium and see that
one has a
fluctuation-dissipation theorem
(FDT) available.  The existence of a FDT allows one to
organize the theory very efficiently.

\subsection{Time-reversal symmetry}

Let us focus on
the time reversal  transformation given by\cite{ABL}
\be
\tau R_{i }(t)=R_{i }(-t)
\ee
\be
\tau\hat{R}_{i }(t)=-\hat{R}_{i}(-t)
+i \beta F_{i}(-t).
\ee
How does the action
\be
A=\int_{-\infty}^{\infty}dt \sum_{i }\Bigg[
[\hat{R}_{i}(t)
 D\beta^{-1}\hat{R}_{i}(t)]
+i\hat{R}_{i}(t)
\left[\frac{\partial}{\partial t}
R_{i}(t)-DF_{i}( R )\right]\Bigg]+A_{J}
\ee
 change under the transformation? We have
\be
A'=\tau A
\nonumber
\ee
\be
=\int_{-\infty}^{\infty}dt\sum_{i} \Bigg[
[-\hat{R}_{i }(-t)+i \beta F_{i }(-t)]
D\beta^{-1}[-\hat{R}_{i}(-t)+i \beta F_{i}(-t)]
\nonumber
\ee
\be
+i[-\hat{R}_{i}(-t)
-i \beta  F_{i}(-t)]
\left[\frac{\partial}{\partial t}
R_{i}(-t)-DF_{i}(\tau R )\right]\Bigg]
+\tau A_{J}.
\ee
We need to look at the various terms.  Consider first
\be
\int_{-\infty}^{\infty}dt\sum_{i} \Bigg[
-i \beta  F_{i}(-t)]
\left[\frac{\partial}{\partial t}
R_{i}(-t)\right]\Bigg]
\nonumber
\ee
\be
=\int_{-\infty}^{\infty}dt
\sum_{i }(- i \beta )(-1) \frac{\delta U(-t)}{\delta R_{i }(-t)}
i\frac{\partial}{\partial t}R_{i }(-t)
\nonumber
\ee
\be
=\int_{-\infty}^{\infty}dt\sum_{i}
(-\beta )
\frac{\partial}{\partial (-t)}U(-t)
=\beta \int_{\infty}^{-\infty}ds
\frac{\partial}{\partial s}U(s)
=0.
\nonumber
\ee
Next we collect the two force terms
\be
\int_{-\infty}^{\infty}dt\sum_{i}
\beta \left[(-1)F_{i}(-t)F_{i}(-t)+F_{i}(-t)F_{i}(-t)\right]=0
\ee
Putting this together we have
\be
A'=\int_{-\infty}^{\infty}dt\sum_{i} \Bigg[
[-\hat{R}_{i }(-t)]
D\beta^{-1}[-\hat{R}_{i}(-t)+2i \beta F_{i}(-t)]
\nonumber
\ee
\be
+i[-\hat{R}_{i}(-t)
\left[\frac{\partial}{\partial t}
R_{i}(-t)-DF_{i}(\tau R )\right]\Bigg]
+\tau A_{J}
\ee
where since $A_{J}$ is a function of the density, $\tau A_{J}=A_{J}$.
Then combining the terms linear in $F$ and letting $t\rightarrow -t$
we have the invariance principle
$A'=A$.

\subsection{Application of the invariance principle to averages of fields}

The first application of the invariance is to correlation functions.
For the $n$-point density correlation function, we have, from its invariance
under $\tau$,
\be
G_{\rho\rho\ldots\rho}(1,2,\ldots,n)
=G_{\rho\rho\ldots\rho}(\tilde{1},\tilde{2},\ldots,\tilde{n})
\ee
where $\tilde{\ell}=(x_{\ell},-t_{\ell})$.

Consider the two-point response function
\be
G_{\rho B}(12)=\langle \rho (1) B(2)\rangle
\nonumber
\ee
\be
=\langle \rho (1) B_{0}(2)\rangle
+\langle \rho (1) B_{J}(2)\rangle
\ee
where the particle density is given by
\be
\rho (1)=\sum_{i=1}^{N}\delta [x_{1}-R_{i}(t_{1})].
\ee
The response field is the sum of
\be
B_{0}(2)=D\sum_{i=1}^{N}i\hat{R}_{i}(t_{2})\cdot\nabla_{x_{2}}
\delta [x_{2}-R_{i}(t_{2})]
\ee
and
\be
B_{J}(2)=\theta(0)D\nabla^{2}_{x_{2}}\rho (2).
\ee
Applying $\tau$ to $B_{0}$
\be
\tau B_{0}(2)=
D\sum_{i=1}^{N}[-\hat{R}_{i}(-t_{2})+i\beta F_{i}(-t_{2})]
\cdot\nabla_{x_{2}}
i\delta [x_{2}-R_{i}(-t_{2})]
\nonumber
\ee
\be
=-B_{0}(\tilde{2})-\beta D\nabla_{x_{2}}F(\tilde{2}).
\ee
We can then write $\tilde{2}=(x_{2},-t_{2})$ and
\be
\tau B(2)
=-B(\tilde{2})+2B_{J}(\tilde{2})-\beta D\nabla_{x_{2}}F(\tilde{2}).
\ee
Then we have on applying the invariance principle
\be
G_{\rho B}(12)=
-G_{\rho B}(\tilde{1},\tilde{2})
+2\theta (0)D\nabla_{x_{2}}^{2}
G_{\rho \rho}(\tilde{1},\tilde{2})
-\beta D\nabla_{x_{2}}G_{\rho F}(\tilde{1},\tilde{2}).
\label{eq:57}
\ee

We now need an independent expression for the density-force density
correlation function.  As a first step construct an operator
$\hat{\cal O}_{B_{0}}(2)$ that satisfies
\be
\hat{\cal O}_{B_{0}}(2)\int_{-\infty}^{\infty}dt\sum_{j=1}^{N}
\hat{R}_{j}(t)\bar{D}\hat{R}_{j}(t)=B_{0}(2).
\ee
It is easy to see that
\be
\hat{\cal O}_{B_{0}}(2)=\sum_{j=1}^{N}\frac{i\beta}{2}\nabla_{x_{2}}\cdot
\frac{\delta}{\delta\hat{R}_{j}(t_{2})}
\delta[x_{2}-R_{j}(t_{2})].
\ee
We now use the identity
\be
Tr f(R)\hat{\cal O}_{B_{0}}(2)e^{-A}=0
\ee
or
\be
\langle  f(R)\hat{\cal O}_{B_{0}}(2)A\rangle =0.
\ee
Letting the operator act on $A$, we have
\be
\hat{\cal O}_{B_{0}}(2)A=B_{0}(2)-\frac{\beta}{2}
\frac{\partial}{\partial t_{2}}\rho (2)+
\frac{\beta D}{2}\nabla_{x_{2}}F(2)
\ee
\be
=B(2)+\frac{\beta}{2}
\left(\frac{-\partial}{\partial t_{2}}-\bar{D}\nabla_{x_{2}}^{2}\right)\rho (2)+
\frac{\beta D}{2}\nabla_{x_{2}}F(2).
\ee
Multiplying by the density and averaging, we have
\be
\bigg\langle \rho(1) \left[B(2)+\frac{\beta}{2}
\left(\frac{-\partial}{\partial t_{2}}-\bar{D}\nabla_{x_{2}}^{2}\right)\rho (2)+
\frac{\beta D}{2}\nabla_{x_{2}}F(2)\right]\bigg\rangle =0
\ee
or
\be
G_{\rho B}(12)+\frac{\beta}{2}\bigg(\frac{-\partial}{\partial t_2}-\bar{D}\nabla^2_{x_2}\bigg)G_{\rho\rho}(12)
+\frac{\beta}{2}\nabla_{x_2}G_{\rho F}(12) = 0.
\label{eq:65}
\ee

Eliminating $G_{\rho F}$ between Eqs.(\ref{eq:57})
and (\ref{eq:65}) we have
\be
G_{\rho  B}(12)=-G_{\rho  B}(\tilde{1},\tilde{2})+
2\theta (0)D\nabla_{x_{2}}^{2}
G_{\rho  \rho}(\tilde{1},\tilde{2})
+2G_{\rho  B}(\tilde{1},\tilde{2})
\nonumber
\ee
\be
-\beta\left(\frac{\partial}{\partial t_{2}}+\bar{D}\nabla_{x_{2}}^{2}\right)
G_{\rho  \rho}(\tilde{1},\tilde{2})
\nonumber
\ee
\be
=G_{\rho  B}(\tilde{1},\tilde{2})
-\beta\frac{\partial}{\partial t_{2}}
G_{\rho  \rho}(\tilde{1},\tilde{2}).
\ee
Since $G_{\rho B}$ is retarded we can multiply by
$\theta (t_{1}-t_{2})$ and obtain the crucial FDT
\be
G_{\rho  B}(12)=
-\theta (t_{1}-t_{2})
\beta\frac{\partial}{\partial t_{2}}
G_{\rho  \rho}(\tilde{1},\tilde{2})
\nonumber
\ee
\be
=\theta (t_{1}-t_{2})
\beta\frac{\partial}{\partial t_{1}}
G_{\rho  \rho}(1,2)
\label{eq:70}.
\ee

\subsection{FDT: Fourier Transform}

The time Fourier transform of Eq.(\ref{eq:70}) is given by
\be
G_{\rho B}(q, \omega )=\int_{-\infty}^{\infty}
dte^{i\omega\cdot t}\beta\theta (t)
\frac{\partial}{\partial t}G_{\rho\rho}(q,t)
\nonumber
\ee
\be
=\int_{0}^{\infty}dt ~e^{i\omega\cdot t}\beta
\frac{\partial}{\partial t}\int_{-\infty}^{\infty}
\frac{d\bar{\omega}}{2\pi }e^{-i\bar{\omega}t}G_{\rho\rho}(q,\bar{\omega})
\nonumber
\ee
\be
=\int_{-\infty}^{\infty}
\frac{d\bar{\omega}}{2\pi }(-i\bar{\omega})
G_{\rho\rho}(q,\bar{\omega})
\int_{0}^{\infty} ~dte^{i(\omega\cdot-\bar{\omega}) t}
\nonumber
\ee
\be
=\int_{-\infty}^{\infty}
\frac{d\bar{\omega}}{2\pi }(-i\bar{\omega})
G_{\rho\rho}(q,\bar{\omega})
\frac{i}{\omega-\bar{\omega}+i\eta}.
\ee
So we have
\be
G_{\rho B}(q, \omega )=
\int_{-\infty}^{\infty}
\frac{d\bar{\omega}}{2\pi }(\beta\bar{\omega})
\frac{G_{\rho\rho}(q,\bar{\omega})}{\omega-\bar{\omega}+i\eta}.
\label{eq:75}
\ee
Taking the imaginary part
\be
Im~
G_{\rho B}(q, \omega )=
\int_{-\infty}^{\infty}
\frac{d\bar{\omega}}{2\pi }(\beta\bar{\omega})
Im ~\frac{G_{\rho\rho}(q,\bar{\omega})}{\omega-\bar{\omega}+i\eta}
\nonumber
\ee
\be
=-\frac{\beta\omega}{2}G_{\rho\rho}(q,\omega ).
\label{eq:76}
\ee
It is useful to
check the FDT
in the noninteracting limit.

\subsection{FDT and vertex functions}

Dyson's equation,
\be
\sum_{\mu}\Gamma_{\alpha\mu}G_{\mu \beta}=\delta_{\alpha\beta},
\label{eq:83}
\ee
is a matrix equation relating matrix elements for the cumulants
to the matrix elements of the vertex functions.  Using
$G_{BB}=0$ and $\Gamma_{\rho\rho}=0$ we have
\be
G_{\rho B}=\frac{1}{\Gamma_{B\rho}},
\ee
\be
G_{B \rho }=\frac{1}{\Gamma_{\rho B}},
\ee
and
\be
G_{\rho\rho}=
-\frac{1}{\Gamma_{B\rho}}\Gamma_{BB}
\frac{1}{\Gamma_{\rho B}}
\label{eq:74}.
\ee
Since $G_{\rho B}=G^{*}_{B\rho }$ and $\Gamma_{B\rho}=\Gamma^{*}_{B\rho}$, we have
\be
G_{\rho\rho}=
-\frac{\Gamma_{BB}}{|\Gamma_{B\rho}|^{2}}
\ee
and since $G_{\rho\rho}$ is positive, we have
\be
\Gamma_{BB}< 0.
\ee
We can now use the FDT in the following way.  Taking the imaginary part
of $G_{\rho B}$,
\be
Im ~G_{\rho B}=
-Im \frac{\Gamma_{\rho B}}{|\Gamma_{B\rho}|^{2}}
\nonumber
\ee
\be
=-\frac{\beta\omega}{2}G_{\rho\rho},
\ee
and using Eq.(\ref{eq:74}) for $G_{\rho\rho}$, we can cancel the positive
denominators and find
\be
Im ~ \Gamma_{\rho B}=\frac{\beta \omega}{2}\Gamma_{BB}
\ee
or
\be
Im ~ \Gamma_{B\rho }=-\frac{\beta \omega}{2}\Gamma_{BB}.
\ee

\subsection{Kinetic equation and the FDT}

Starting with the $B\rho$ component of Dyson's equation in
$q, t$ space, we have
\be
\int d\bar{t}~\Gamma_{B\rho}(q,t-\bar{t})G_{\rho\rho}(q,\bar{t}-t')
+\int d\bar{t}~\Gamma_{BB}(q,t-\bar{t})G_{B\rho}(\bar{t}-t')=0.
\label{eq:80}
\ee
Let us now write
\be
\Gamma_{\alpha\beta}(q,t)=
\gamma^{(1)}_{\alpha\beta}(q,t)-
\Sigma_{\alpha\beta}(q,t)
\ee
where $\gamma^{(1)}_{\alpha\beta}(q,t)$ includes the zeroth and first order
contributions in perturbation theory and the self-energy $\Sigma$ starts
at second order.  Suppressing the wavenumber label, we can rewrite
Eq.(\ref{eq:80})
in the form
\be
\gamma_{B\rho}^{(1)}(t)G_{\rho\rho}(t,t')
+\gamma_{BB}^{(1)}(t)G_{B\rho}(t,t')
=\Psi (t,t'),
\ee
where
\be
\Psi (t,t')=
\int_{-\infty}^{t}ds \Sigma_{B\rho}(t-s)G_{\rho\rho}(s-t')
+\int_{-\infty}^{t'}ds
\Sigma_{BB}(t-s)G_{B\rho}(s-t')
\ee
using the fact that $\Sigma_{B\rho}(t-s)\sim \theta (t-s)$
and $G_{B\rho}(s-t')\sim \theta (t'-s)$.  We then use the
fluctuation-dissipation theorems
\be
\Sigma_{B\rho}(t-s)=\theta (t-s)\beta\frac{\partial}{\partial t}
\Sigma_{BB}(t-s)
\ee
\be
G_{B\rho}(s-t')=\theta (t'-s)\beta\frac{\partial}{\partial t'}
G_{\rho\rho}(t'-s)
\ee
to obtain
\be
\Psi (t,t')=
-\int_{-\infty}^{t}ds \left[\beta\frac{\partial}{\partial s}
\Sigma_{BB}(t-s)\right]
G_{\rho\rho}(s-t')
-\int_{-\infty}^{t'}ds
\Sigma_{BB}(t-s)
\beta\frac{\partial}{\partial s}
G_{\rho\rho}(t'-s).
\ee
If we integrate the first integral by parts, we have
\be
\Psi (t,t')=- \beta\Sigma_{BB}(0)G_{\rho\rho}(t-t')
+\beta\int_{-\infty}^{t}ds\Sigma_{BB}(t-s)
\frac{\partial}{\partial s}G_{\rho\rho}(s-t')
\nonumber
\ee
\be
-\int_{-\infty}^{t'}ds
\Sigma_{BB}(t-s)
\frac{\partial}{\partial s}
G_{\rho\rho}(t'-s)
\nonumber
\ee
\be
=-\Sigma_{BB}(0)G_{\rho\rho}(t-t')
+\int_{t'}^{t}ds \Sigma_{BB}(t-s)
\frac{\partial}{\partial s}
G_{\rho\rho}(t'-s)
\ee
where we assume $t > t'$.
We then have the kinetic equation
\be
\beta\gamma_{B\rho}^{(1)}(t)G_{\rho\rho}(t,t')
=-\beta^2\Sigma_{BB}(0)G_{\rho\rho}(t-t')
+\beta^2\int_{t'}^{t}ds \Sigma_{BB}(t-s)
\frac{\partial}{\partial s}
G_{\rho\rho}(t'-s).
\label{eq:101}
\ee
We see that our dynamical problem is now in the form of a
memory function equation and the dynamic part of the memory function
is given by the self-energy $\Sigma_{BB}$.  With some additional work, one can show the equal-time quantity
\be
\beta^2\rho_0\Sigma_{BB}(q,t=0)=S^{-1}(q)-[1+\rho_0\beta V(0)],
\ee
where $V(q)$ is the Fourier transform of the potential. We will discuss these results in great detail elsewhere.

\subsection{Static Implications}

The integral form of the FDT tells us that in the small $\omega$
limit
\be
G_{\rho B}(q, 0 )=
\int_{-\infty}^{\infty}
\frac{d\bar{\omega}}{2\pi }(-\beta )
G_{\rho\rho}(q,\bar{\omega})
\ee
\be
=-\beta S(q)
\ee
where $S(q)$ is the static structure factor.
One can  also write this as
\be
S(q)=-\frac{k_{B}T}{\Gamma_{B\rho}(q,0)}.
\ee
This result will be extremely useful to us.

\section{Irreducible Vertex Functions and Brownian Gas Cumulants}

\subsection{Irreducible vertex functions}
We list here the fundamental definitions relating the cumulants in the theory to the reducible vertex functions we need. The two-point irreducible vertex $\Gamma_{ij}$ is defined by the Dyson's equation
\be
\Gamma_{ik}G_{kj}=\delta_{ij}.
\label{eq:star}
\ee

The three-point one-particle irreducible vertex is defined by
\be
\Gamma_{ijk}=\frac{\delta\Gamma_{ij}}{\delta G_k}
\ee
which is equivalent to
\be
G_{ijk}=\frac{\delta G_{ij}}{\delta H_k} = -G_{iu}G_{jv}G_{kw}\Gamma_{uvw}
\ee
which can be rewritten using Eq. (\ref{eq:star}) as
\be
\Gamma_{ijk}=-\Gamma_{iu}\Gamma_{jv}\Gamma_{kw}G_{uvw}.
\ee
The case of four-point cumulants and irreducible vertex functions is discussed elsewhere\cite{four-point}.

\subsection{Noninteracting Brownian particle cumulants:
wavenumber and time regime}

In FTSPD we derived a compact expression for the $n$-point cumulants
for the fields $\phi =(\rho, B)$.  In the time wavenumber regime, we have
\be
G_{\alpha_{1},\alpha_{2},\ldots ,\alpha_{n}}^{(0)}(1,2,\ldots, n)
=\rho_{0}\delta (q_{1}+q_{2}+\ldots +q_{n})b_{1}b_{2}\ldots b_{n}
e^{ N_{n}}
\ee
where $b_{j}=b_{\alpha_{j}}(q_{j},t_{j})$
with
\be
b_{\rho}(q_{j},t_{j})=1,
\ee
\be
b_{B}(q_{j},t_{j})=\beta \sum_{k=1\neq j}^{n}K_{jk}\theta (t_{k}-t_{j}),
\ee
\be
N_{n}=\frac{1}{2}\sum_{i,j=1}^{n}K_{ij}|t_{i}-t_{j}|,
\ee
and
\be
K_{ij}=\bar{D}q_{i}\cdot q_{j}.
\ee
This expression is manifestly translationally invariant in space and time.

For our purposes, we need these cumulants Fourier transformed over time:
\be
G_{\alpha_{1},\alpha_{2},\ldots ,\alpha_{n}}^{(0)}(1,2,\ldots, n)
=\rho_{0}\delta (q_{1}+q_{2}+\ldots +q_{n})
\nonumber
\ee
\be
\times
\int dt_{1}dt_{2}\ldots dt_{n} b_{1}b_{2}\ldots b_{n}
e^{i\sum_{k=1}^{n}\omega_{k}t_{k}}e^{N_{n}},
\ee
where we have not found it confusing to use the same symbol for the Fourier
transform.
In this paper we assume the system is in an equilibrium state for all time.

\subsection{Time Fourier transform}
We discuss the method of taking these time Fourier transforms elsewhere\cite{four-point}. Here, we simply list the needed results. We will need the zeroth order two-point cumulants
\be
G_{\rho\rho}^{(0)}(12)=2\kappa_{1}\rho_{0}G_{1}G_{1}^{*}\delta (1+2),
\ee
\be
G^{(0)}_{B\rho}(12)=-\beta \kappa_{1}\rho_{0}G_{1}^{*}\delta (1+2),
\ee
\be
G^{(0)}_{\rho B}(12)=-\beta \kappa_{1}\rho_{0}G_{1}\delta (1+2),
\ee
and
\be
G^{(0)}_{B B}(12)=0,
\ee
where
\be
\kappa_i = K_{ii} = \bar{D}q_i^2,
\ee
\be
G_i^{-1}=-i\omega_i+\kappa_i,
\ee
and
\be
\delta(1+2) = (2\pi)^d\delta(q_1+q_2) 2\pi\delta(\omega_1+\omega_2).
\ee

\subsection{Two-point vertices}
We also need the zeroth order two-point irreducible vertices defined by
\be
\gamma_{ik}^{(0)}G_{kj}^{(0)}=\delta_{ij}.
\ee
We find
\be
\beta\gamma^{(0)}_{\rho\rho}(12)=0,
\ee
\be
\beta\gamma^{(0)}_{B\rho}(12)=-\frac{G_{1}^{-1}}{\rho_{0}\kappa_{1}}\delta(1+2),
\ee
\be
\beta\gamma^{(0)}_{\rho B}(12)=-\frac{G_{1}^{-1,*}}{\rho_{0}\kappa_{1}}\delta(1+2),
\ee
and
\be
\beta^{2}\gamma^{(0)}_{B B}(12)=-\frac{2}{\rho_{0}\kappa_{1}}\delta(1+2).
\ee
These are the key building blocks of the theory.

\subsection{Three-point vertices}
A key role in the second order theory is played by the zeroth order three-point irreducible vertices. These are defined in terms of the three-point cumulants
\be
\gamma_{ijk}^{(0)}(123)=-\gamma_{iu}^{(0)}(1)\gamma_{jv}^{(0)}(2)\gamma_{kw}^{(0)}(3)G_{uvw}(123).
\ee

These can be found to be given by
\be
\gamma_{\rho\rho\rho}^{(0)}(123)=\gamma_{BBB}^{(0)}(123)=0,
\ee
\be
\beta\gamma_{B\rho\rho}^{(0)}(123)=
-\frac{1}{\rho_{0}^{2}}
\left[\bar{K}_{12}G_{2}^{-1,*}
+\bar{K}_{13}G_{3}^{-1,*}\right]
\delta (1+2+3)=\frac{1}{\rho_{0}^{2}}[1-iE_{1}]
\delta (1+2+3),
\ee
and
\be
\beta\gamma_{BB\rho}^{(0)}(123)=
-2\frac{1}{\rho_{0}^{2}}\bar{K}_{12}
\delta (1+2+3)
\ee
where
\be
\bar{K}_{ij}=\frac{K_{ij}}{\kappa_{i}\kappa_{j}},
\ee
\be
\delta(1+2+3)=(2\pi)^d\delta(q_1+q_2+q_3) 2\pi \delta(\omega_1+\omega_2+\omega_3)
\ee
and
\be
E_{1}=\omega_{2}\bar{K}_{12}+\omega_{3}\bar{K}_{13}.
\ee

The other vertices can be constructed using symmetry.

\section{First order result for $K$}

Going to first order in the potential we  have
\be
\Delta W^{(1)}=\sum_{u}F_{u}G_{u}.
\ee
We next need to compute
\be
\frac{\delta}{\delta G_{j}}\Delta W^{(1)}
=F_{j}
\ee
which goes into Eq.(\ref{eq:34}) giving
the result
\be
c_{ij}=\langle \phi_{i}F_{j}\rangle =\sum_{u}{\cal G}_{iu}\sigma_{uj}.
\ee
Putting this result into Eq.(\ref{eq:38}) and using Eq.(\ref{eq:39})
gives the very simple result
\be
K^{(1)}_{ij}=-\sigma_{ij}.
\ee
These results satisfy the FDT in a trivial way:
\be
K_{\rho B}^{(1)}=K_{B\rho }^{(1)}=-\beta V(q)
\ee
and
\be
K_{B B}^{(1)}=-\frac{2}{\beta \omega}Im ~K_{B\rho }^{(1)}=0.
\ee

\section{Full First Order Solution}

In FTSPD we worked out the first order theory concentrating
on the two-point cumulant.  Here we work things out including
the equation of state and one-point averages.
This sets the stage for the second order calculation.

At first order in perturbation theory\cite{FTP}, we have the equation of state
\be
G_{i}=G_{i}^{(0)}+\tilde{T}r \phi_{i}\Delta W^{(1)}+...
\ee
and for the two-point cumulant, the kinetic equation
\be
G_{ij}=G_{ij}^{(0)}+\tilde{T}r \phi_{i}\phi_{j}\Delta W^{(1)}+
G_{i\ell}^{(0)}\sigma_{\ell u}G_{uj}.
\ee
First, we rewrite the first order static term
\be
G_{i}^{(1)}=\tilde{T}r \phi_{i}\Delta W^{(1)}
=\tilde{T}r \phi_{i}F_{u}G_{u}
\nonumber
\ee
\be
=\tilde{T}r \phi_{i}\phi_{\ell}\sigma_{\ell u}G_{u}
=G_{i\ell}^{(0)}\sigma_{\ell u}G_{u}
\ee
and the two-point function
\be
{\cal G}_{ij}^{(1)}=\tilde{T}r \phi_{i}\phi_{j}\Delta W^{(1)}
=\tilde{T}r \phi_{i}\phi_{j}F_{u}G_{u}
\nonumber
\ee
\be
=\tilde{T}r \phi_{i}\phi_{j}\phi_{\ell}\sigma_{\ell u}G_{u}
=G_{ij\ell}^{(0)}\sigma_{\ell u}G_{u}.
\ee
Thus, one has the coupled set of equations
\be
G_{i}=G_{i}^{(0)}+G_{i\ell}^{(0)}\sigma_{\ell u}G_{u}
\label{eq:206}
\ee
and
\be
G_{ij}=G_{ij}^{(0)}+G_{ij\ell}^{(0)}\sigma_{\ell u}G_{u}
+G_{i\ell}^{(0)}\sigma_{\ell u}G_{uj}
\ee
where in the field theory protocol
\be
\sigma_{ij}(k)=V(k)\left[\hat{\rho}_{i}\hat{B}_{j}
+\hat{B}_{i}\hat{\rho}_{j}\right].
\ee
Let us look first at the term with the cubic cumulant
\be
{\cal G}_{ij}^{(1)}
=G_{ij\ell}^{(0)}\sigma_{\ell u}G_{u}.
\ee
We can write
\be
G_{ij\ell}^{(0)}=-G_{iu}^{(0)}G_{jv}^{(0)}G_{\ell w}^{(0)}
\gamma_{uvw}^{(0)}
\ee
and thus
\be
{\cal G}_{ij}^{(1)}=
-G_{iu}^{(0)}G_{jv}^{(0)}G_{\ell w}^{(0)}
\gamma_{uvw}^{(0)}\sigma_{\ell s}G_{s}
\nonumber
\ee
\be
=-G_{iu}^{(0)}G_{jv}^{(0)}\gamma_{uvw}^{(0)}
\left(G_{w}-G_{w}^{(0)}\right)
\ee
where in the last step we used the equation of state, Eq.(\ref{eq:206}).
If we use the useful identity
\be
\gamma_{uvw}^{(0)}G_{w}^{(0)}=-\gamma_{uv}^{(0)}
\ee
and realize that we can write
\be
G_{i}=\frac{\bar{\rho}}{\rho_{0}}G_{i}^{(0)}
\ee
where  $\bar{\rho}=\langle\rho\rangle $, we have
\be
{\cal G}_{ij}^{(1)}=
-G_{iu}^{(0)}G_{jv}^{(0)}[-\gamma_{uv}^{(0)}(\bar{\rho}/\rho_{0}-1)]
\nonumber
\ee
\be
=G_{ij}^{(0)}(\bar{\rho}/\rho_{0}-1).
\ee
The two-point equation becomes
\be
G_{ij}=G_{ij}^{(0)}+G_{ij}^{(0)}(\bar{\rho}/\rho_{0}-1)
+G_{i\ell}^{(0)}\sigma_{\ell u}G_{uj}
\nonumber
\ee
\be
=\frac{\bar{\rho}}{\rho_{0}}G_{ij}^{(0)}
+G_{i\ell}^{(0)}\sigma_{\ell u}G_{uj}.
\label{eq:188}
\ee
Now $\bar{\rho}/\rho_{0}$ is determined by the equation of state.
If we assume the one-point average is of the form
\be
G_{u}(1)=\hat{\rho}_{u}G_{\rho}(q_{1})2\pi \delta (\omega_{1})
\ee
then we have
\be
\hat{\rho}_{i}[G_{\rho}(q_{1})-G_{\rho}^{(0)}(q_{1})]
2\pi \delta (\omega_{1})
=G_{i\ell}^{(0)}(q_{1},\omega_{1})\sigma_{\ell u}(q_{1})
G_{u}(q_{1},\omega_{1})
\nonumber
\ee
\be
=G_{i\ell}^{(0)}(q_{1},0)\sigma_{\ell u}(q_{1})
\hat{\rho}_{u}G_{\rho}(q_{1})2\pi \delta (\omega_{1})
\nonumber
\ee
\be
=G_{i\ell}^{(0)}(q_{1},0)\sigma_{\ell \rho}(q_{1})
G_{\rho}(q_{1})2\pi \delta (\omega_{1})
\nonumber
\ee
\be
=G_{iB}^{(0)}(q_{1},0)V(q_{1})
G_{\rho}(q_{1})2\pi \delta (\omega_{1})
\nonumber
\ee
\be
=\hat{\rho}_{i}
G_{\rho B}^{(0)}(q_{1}, 0)V(q_{1})
G_{\rho}(q_{1})2\pi \delta (\omega_{1})
\nonumber
\ee
\be
=\hat{\rho}_{i}
(-\rho_{0})\beta V(q_{1})
G_{\rho}(q_{1})2\pi \delta (\omega_{1}).
\ee
Canceling common factors gives
\be
G_{\rho}(q)=
G_{\rho}^{(0)}(q)-\rho_{0}\beta V(q)
G_{\rho}(q)
\ee
or
\be
G_{\rho}(q)=\frac{1}{1+\rho_{0}\beta V(q)}G_{\rho}^{(0)}(q).
\ee
In the homogeneous limit
\be
\bar{\rho} =\frac{\rho_{0}}{1+\rho_{0}\beta V(0)}.
\ee

Turn next to the two-point correlations which satisfy
Eq.(\ref{eq:188}).  Taking
the $\rho B$ matrix element,
one gets a closed equation
\be
G_{\rho B}(1)=(\bar{\rho}/\rho_{0})
G_{\rho B}^{(0)}(1)
+G_{\rho B}^{(0)}(1)V(q_{1})
G_{\rho B}(1)
\ee

which can be solved to give
\be
G_{\rho B}(q_{1})=\frac{(\bar{\rho}/\rho_{0})G_{\rho B}^{(0)}(1)}
{1-G_{\rho B}^{(0)}(1)V(q_{1})}
\ee
\be
=-\frac{\beta\bar{\rho} \kappa_{1}}{-i\omega_{1} +\kappa_{1} [1+\tilde{V}(q_{1})]}
\ee
where $\kappa_{1} =\bar{D}q_{1}^{2}$ and
$\tilde{V}(q_{1})=\rho_{0}\beta V(q_{1})$.  One can then solve for
the density-density correlation function
\be
G_{\rho\rho}(1)=(\bar{\rho}/\rho_{0})G_{\rho\rho}^{(0)}(1)+G_{\rho\rho}^{(0)}(1)
V(q_{1}) G_{B\rho}(1)+
G_{\rho B}^{(0)}(1)V(q_{1})G_{\rho\rho}(1)
\ee
or
\be
G_{\rho\rho}(1)=\frac{1}{1-G_{\rho B}^{(0)}(1)V(q_{1})}
\left((\bar{\rho}/\rho_{0})G_{\rho\rho}^{(0)}(1)+G_{\rho\rho}^{(0)}(1)
V(q_{1}) G_{B\rho}(1)\right).
\ee
Putting in the results for the zeroth order correlations and
\be
G_{B\rho }^{(1)}(1)
=-\frac{\bar{\rho} \beta\kappa_{1}}{i\omega_{1} +\kappa_{1} [1+\tilde{V}(q_{1})]}
\ee
leads to the final result
\be
G_{\rho\rho}(1)=\frac{2\bar{\rho} \kappa_{1}}{\omega^{2}_{1}
+\kappa^{2}_{1}[1+\tilde{V}(q_{1})]^{2}}.
\ee

Turning to the statics, we can use the result found earlier
\be
-\beta S(q)=G_{\rho B}(q, 0)=\frac{-\bar{\rho}\beta\kappa}
{\kappa [1+\tilde{V}(q_{1})]}
\ee
or
\be
 S(q)=\frac{\bar{\rho}}
{1+\tilde{V}(q_{1})}.
\label{eq:227a}
\ee
Comparing Eq.(\ref{eq:227a}) with the
static Ornstein-Zernike\cite{OZR}
relation we can  identify the effective
interaction
\be
V_{EFF}(q)=-\beta^{-1}c_{D}(q)
\ee
where $c_{D}(q)$ is the physical direct correlation function which is assumed
to be known by other means.  We can, for example, assume that
$c_{D}(q)$ is given in the Percus-Yevick approximation for hard
spheres\cite{PY}.

\section{Self-consistent Collective Self-energy at Second Order }

The evaluation of the collective part of the  self-energy,
 $K$, at second order begins with determining
$f_{j}$, defined by
\be
f_{j}=\frac{\delta}{\delta G_{j}}\Delta W
\ee
at second order.  We have
\be
\Delta W^{(2)}=\frac{1}{2}F_{u}F_{v}G_{uv}
\ee
and
\be
f_{j}^{(2)}=\frac{\delta \Delta W^{(2)}}{\delta G_{j}}
=\frac{1}{2}F_{u}F_{v}\frac{\delta G_{uv}}{\delta G_{j}}
\label{eq:203}.
\ee
We then have the standard functional manipulations,
\be
\frac{\delta}{\delta G_{k}}G_{ij}
=-\sum_{uv}G_{iu}\frac{\delta}{\delta G_{k}}
G^{-1}_{uv}G_{vj}
\nonumber
\ee
\be
=-\sum_{uv}G_{iu}G_{jv}\Gamma^{(3)}_{uvk}
\ee
where we introduce the three-point vertex
\be
\Gamma_{ijk}=\frac{\delta}{\delta G_{k}}\Gamma_{ij}.
\ee
Putting this back into Eq.(\ref{eq:203}) gives
\be
f_{j}^{(2)}=\frac{\delta \Delta W^{(2)}}{\delta G_{j}}
=-\frac{1}{2}F_{\ell}F_{n}
G_{\ell u}G_{nv}\Gamma_{uvj}
\ee
This in turn goes into Eq.(\ref{eq:34}) and
\be
c_{ij}^{(2)}=-\frac{1}{2}\tilde{T}r \phi_{i}
e^{\Delta W}F_{\ell}F_{n}
G_{\ell u}G_{nv}\Gamma_{uvj}
\nonumber
\ee
\be
=-\frac{1}{2}{\cal G}_{i\ell s}
\sigma_{\ell n}\sigma_{sm}
G_{nu}G_{mv}\Gamma_{uvj}
\nonumber
\ee
where
\be
{\cal G}_{i\ell s}=\tilde{T}r~ \phi_{i}\phi_{\ell}
\phi_{s}e^{\Delta W}
\ee
is a three-point self-correlation.  Things can be written
more symmetrically if we introduce the three-point self-vertex
$\gamma_{ijk}$ via
\be
{\cal G}_{i\ell s}=
-{\cal G}_{iu}
{\cal G}_{\ell v}
{\cal G}_{sw}\gamma_{uvw}.
\ee
Then, the collective part of the two-point vertex is given at second order by
\be
K_{ij}=-\frac{1}{2}\gamma_{in}
{\cal G}_{nn'}{\cal G}_{\ell\ell'}
{\cal G}_{ss'}\gamma_{n'\ell's'}
\sigma_{\ell u}\sigma_{sv}
G_{ur}G_{vq}\Gamma_{jrq}
\nonumber
\ee
\be
=-\frac{1}{2}
\gamma_{i\ell's'}
{\cal G}_{\ell\ell'}
{\cal G}_{ss'}
\sigma_{\ell u}\sigma_{sv}
G_{ur}G_{vq}\Gamma_{rqj}
\nonumber
\ee
\be
=-\frac{1}{2}
\gamma_{iuv}
\bar{G}_{uw}\bar{G}_{vz}\Gamma_{wzj}
\nonumber
\ee
where we have the self-dressed propagators
\be
\bar{G}_{ur}={\cal G}_{uv}\sigma_{vs}G_{sr}.
\ee
We see already the structure for making vertex corrections.
At lowest order in the interaction we have
\be
\Gamma^{(0)}_{ijk}=\gamma^{(0)}_{ijk}
\ee
and we have the nontrivial approximation for the second order
contribution to the collective part of the two point vertex
(self-energy) given by
\be
K_{ij}^{(2)}=-\frac{1}{2}\gamma^{(0)}_{iuv}
\bar{G}_{ur}\bar{G}_{vq}\gamma^{(0)}_{krq}.
\label{eq:238}
\ee
Clearly $K$ can be constructed to be symmetric and therefore
$\bar{G}$ can be constructed to be symmetric:
\be
\bar{G}_{ij}=\frac{1}{2}\left({\cal G}_{iv}\sigma_{vs}G_{sj}
+G_{is}\sigma_{sv}
{\cal G}_{vj}\right).
\ee
Equation (\ref{eq:238})
is the key result to be analyzed carefully.  We have a
collective contribution to the two-point vertex which has the
following desired properties:

(i)  It is quadratic functional of the physical
density-density correlation function.

(ii)  We will show that it satisfies the FDT.

(iii)  In the short-time limit, it gives the correct static
contribution to the static structure factor at second order
in the effective potential.

(iv)  It determines the phase diagram for ENE transitions.

In addition to the collective contribution to the
two-point vertex, there is a single-particle contribution.  This part
of the self-energies is very interesting and will be fully
treated in a companion paper.

\section{Self-Dressing Propagators}

\subsection{Second order self-energies}

It appears not to be a coincidence that the matrix propagator
\be
\bar{G}_{ij}=\frac{1}{2}\sum_{uv}\left[G_{iu}^{(0)}\sigma_{uv}G_{vj}
+G_{iu}\sigma_{uv}G_{vj}^{(0)}\right]
\ee
appears in the one-loop expression for the collective part of
the second order self-energy.  In treating the single-particle
contribution to  the second order self-energy we find
another quantity
\be
\tilde{G}_{ij}=\sum_{uvpq}G_{iu}^{(0)}\sigma_{uv}G_{vp}
\sigma_{pq}G_{qj}^{(0)},
\ee
which appears in the theory.  These {\it complications} turn out to be welcome
since $\bar{G}$ and $\tilde{G}$ can be treated as effective matrix
propagators which satisfy the FDT themselves.  They also approach
zero faster than $G$ as $q$ and $\omega$ go to infinity,
thus ensuring convergence of integrals in perturbation theory.
We expect additional similar quantities to appear at higher order
in perturbation theory.

\subsection{$\bar{G}$-frequency regime}

Consider the effective matrix propagator in Fourier space[($1 = (q_{1},\omega_{1})$] given by
\be
\bar{G}_{\alpha\beta}(1)
=\frac{1}{2}\sum_{\mu\nu}\left[G_{\alpha \mu}^{(0)}(1)\sigma_{\mu\nu}(1)
G_{\nu\beta}(1)
+G_{\alpha \mu}(1)\sigma_{\mu\nu}(1)
G_{\nu\beta}^{(0)}(1)\right].
\label{eq:242}
\ee
We assume here that we are working in the field-theory protocol where
\be
\sigma_{\mu\nu}(k)=V(k)(\hat{\rho}_{\mu}\hat{B}_{\nu}
+\hat{\rho}_{\nu}\hat{B}_{\mu}).
\ee
Taking components of Eq.(\ref{eq:242}) we see that the symmetrization does not
influence the response contributions:
\be
\bar{G}_{\rho B}(1)
=\frac{1}{2}\sum_{\mu\nu}\left[G_{\rho \mu}^{(0)}(1)\sigma_{\mu\nu}(1)
G_{\nu B}(1)
+G_{\rho \mu}(1)\sigma_{\mu\nu}(1)
G_{\nu B}^{(0)}(1)\right]
\nonumber
\ee
\be
=\frac{1}{2}\sum_{\mu}\left[G_{\rho \mu}^{(0)}(1)\sigma_{\mu \rho}(1)
G_{\rho B}(1)
+G_{\rho \mu}(1)\sigma_{\mu \rho}(1)
G_{\rho  B}^{(0)}(1)\right]
\nonumber
\ee
\be
=\frac{1}{2}\left[G_{\rho B}^{(0)}(1)V(1)
G_{\rho B}(1)
+G_{\rho B}(1)V(1)
G_{\rho  B}^{(0)}(1)\right]
\nonumber
\ee
\be
=G_{\rho B}^{(0)}(1)V(1)G_{\rho B}(1).
\ee
$\bar{G}_{\rho B}$ and $G_{\rho B}$
have the same analytic structure; they are analytic
in the upper half-plane. Next,
\be
\bar{G}_{B B} =0,
\ee
while for the density-density component,
\be
\bar{G}_{\rho\rho}(1)
=\frac{1}{2}\sum_{\mu\nu}\left[G_{\rho\mu}^{(0)}(1)\sigma_{\mu\nu}(1)
G_{\nu\rho}(1)
+G_{\rho \mu}(1)\sigma_{\mu\nu}(1)
G_{\nu\rho}^{(0)}(1)\right]
\nonumber
\ee
\be
=\frac{1}{2}V(1)\left[G_{\rho B}^{(0)}(1)G_{\rho\rho}(1)
+G_{\rho \rho}^{(0)}(1)G_{B\rho}(1)
+G_{\rho B}(1)G_{\rho\rho}^{(0)}(1)
+G_{\rho \rho}(1)G_{B\rho}^{(0)}(1)\right]
\nonumber
\ee
\be
=\frac{1}{2}V(1)\left[G_{\rho \rho}^{(0)}(1)
\left(G_{B\rho}(1)+G_{\rho B}(1)\right)
+G_{\rho \rho}(1)\left(G_{B\rho}^{(0)}(1)+G_{\rho B}^{(0)}(1)\right)
\right]
\nonumber
\ee
\be
=G_{\rho \rho }^{(0)}(1)V(1) Re G_{\rho B}(1)
+G_{\rho \rho }(1)V(1) Re G_{\rho B}^{(0)}(1)
\ee
which is real but not necessarily positive.
It is crucial to realize that $\bar{G}$ itself satisfies the FDT if
$G$ and $G^{(0)}$ satisfy the FDT.

The proof is as follows. Starting with
\be
\bar{G}_{\rho B}(1) =G_{\rho B}^{(0)}(1) V(1) G_{\rho B}(1)
\ee
and assuming $V(1)$ is real, we take the imaginary part,
\be
Im ~\bar{G}_{\rho B}(1)  =
Im ~G_{\rho B}^{(0)}(1) V(1)  Re ~G_{\rho B}(1)
+Re ~G_{\rho B}^{(0)}(1) V(1)  Im ~G_{\rho B}(1).
\ee
Multiplying by $-2/(\beta\omega_{1})$, we have
\be
-\frac{2}{\beta\omega_{1}}~Im \bar{G}_{\rho B}(1) =
-\frac{2}{\beta\omega_{1}}\left(
Im ~ G_{\rho B}^{(0)}(1)V(1) Re ~G_{\rho B}(1)
+Re ~G_{\rho B}^{(0)}(1)V(1) Im ~G_{\rho B}(1)\right)
\nonumber
\ee
\be
=G_{\rho \rho}^{(0)}(1)V(1) Re ~G_{\rho B}(1)
+Re ~G_{\rho B}^{(0)}(1)V(1)  G_{\rho \rho}(1)
\nonumber
\ee
\be
=\bar{G}_{\rho\rho}(1).
\ee
Thus $\bar{G}$
satisfies the FDT
\be
\bar{G}_{\rho\rho}(\omega)
=-\frac{2}{\beta\omega}Im \bar{G}_{\rho B} (\omega)
\ee
and we can write
\be
\bar{G}_{\rho B}(\omega )=\int\frac{d\bar{\omega}}{2\pi}
\frac{\beta\bar{\omega}\bar{G}_{\rho\rho}(\bar{\omega})}
{\omega-\bar{\omega}+i\eta}
\ee
and
\be
\bar{G}_{B\rho }(\omega )=\int\frac{d\bar{\omega}}{2\pi}
\frac{\beta\bar{\omega}\bar{G}_{\rho\rho}(\bar{\omega})}
{\omega-\bar{\omega}-i\eta}.
\ee

\subsection{$\bar{G}$-time domain}

In the time domain we have the convolution
\be
\bar{G}_{\rho\rho}(t)
=\frac{1}{2}V(k)\int_{-\infty}^{\infty}ds
\left(G_{B\rho}(t-s)+G_{\rho B}(t-s)\right)
G_{\rho \rho}^{(0)}(s)
\nonumber
\ee
\be
+\frac{1}{2}V(k)\int_{-\infty}^{\infty}ds
G_{\rho \rho}(t-s)
\left(G_{B\rho}^{(0)}(s)+G_{\rho B}^{(0)}(s)\right)
\nonumber.
\ee
This together with the fluctuation-dissipation theorem leads to the short-time results
\be
\bar{G}_{\rho\rho}(0)=-\beta V(k)\rho_{0}S(k)
\nonumber
\ee
\be
\dot{\bar{G}}_{\rho\rho}(0)=0
\nonumber
\ee
\be
\ddot{\bar{G}}_{\rho\rho}(0)=\beta V(k)\dot{\bar{G}}_{\rho\rho}(0)
\dot{\bar{G}}_{\rho\rho}^{(0)}(0)
=0.
\ee

\section{Second Order Collective Self-energy and the FDT}

We have the crucial one-loop contribution to the self-energy
\be
\Gamma^{(2,C)}_{\alpha\beta}(12)=-\frac{1}{2}\int d3d4d5d6
\gamma_{\alpha\mu\nu}^{(0)}(134)
\nonumber
\bar{G}_{\mu\sigma}(35)\bar{G}_{\nu q}(46)
\gamma_{\beta \sigma q}^{(0)}(256).
\nonumber
\ee
In terms of Fourier transforms in space and time we find
\be
\Gamma^{(2,C)}_{\alpha\beta}(p_{1},p_{2})=(2\pi)^{d+1}\delta (p_{1}+p_{2})
\Gamma^{(2,C)}_{\alpha\beta}(p_{1})
\ee
and
\be
\Gamma^{(2,C)}_{\alpha\beta}(-p_{1})=-\int dp_{3}dp_{4}
\frac{1}{2}\gamma^{(0)*}_{\alpha\mu\nu}(134)\delta (p_{1}+p_{3}+p_{4})
\nonumber
\ee
\be
\times
\bar{G}_{\mu\sigma}(3)\bar{G}_{\nu q}(4)
\gamma_{\beta \sigma q}^{(0)}(1,3,4)
\label{eq:347a}
\ee
where we have introduced the notation
\be
\int dp_{3} =\int \frac{d\omega_{3}}{2\pi}
\int \frac{d^{d}k_{3}}{(2\pi )^{d}}.
\ee
$\bar{G}$ is a correlation function that satisfies
the fluctuation-dissipation theorem in the form
\be
\bar{G}_{\rho B}(\omega )=\int\frac{d\bar{\omega}}{2\pi}
\frac{\beta\bar{\omega}\bar{G}_{\rho\rho}(\bar{\omega})}
{\omega-\bar{\omega}+i\eta}
\ee
and
\be
\bar{G}_{B\rho }(\omega )=\int\frac{d\bar{\omega}}{2\pi}
\frac{\beta\bar{\omega}\bar{G}_{\rho\rho}(\bar{\omega})}
{\omega-\bar{\omega}-i\eta}.
\ee

\subsection{$\Gamma_{B\rho}$}

We can break Eq.(\ref{eq:347a}) up into components and associate
a set of one-loop contributions which differ by different vertices
and propagators. A number of self-energy contributions vanish
due to causality and for $\Gamma_{B\rho}$ we obtain three
contributions. (To simplify the notation we suppress the
superscript $0$ on the three-point vertices.) Thus
\be
\Gamma_{B\rho}^{(2,C,1)}(-p_{1})=-\frac{1}{2}\int dp_{3}dp_{4}
\gamma_{B\rho\rho}^{*}(134)
\delta (p_{1}+p_{3}+p_{4})\bar{G}_{\rho B}(3)\bar{G}_{\rho B}(4)
\gamma_{\rho BB}(134),
\ee
\be
\Gamma_{B\rho}^{(2,C,2)}(-p_{1})=-\frac{1}{2}\int dp_{3}dp_{4}
\gamma_{B\rho\rho}^{*}(134)
\delta (p_{1}+p_{3}+p_{4})\bar{G}_{\rho B}(3)\bar{G}_{\rho \rho}(4)
\gamma_{\rho B\rho}(134),
\ee
and
\be
\Gamma_{B\rho}^{(2,C,3)}(-p_{1})=-\frac{1}{2}\int dp_{3}dp_{4}
\gamma_{B\rho\rho}^{*}(134)
\delta (p_{1}+p_{3}+p_{4})\bar{G}_{\rho B}(4)\bar{G}_{\rho \rho}(3)
\gamma_{\rho\rho B}(134).
\ee
We want to write these expressions in terms of $G_{\rho\rho}$ and we must
exhibit care in treating the frequency integrals.
Noting that $\beta^{2}\gamma_{\rho B B}(134)=-2\bar{K}_{34}$, we have
\be
\beta\Gamma_{B\rho}^{(2,C,1)}(-p_{1})=-\frac{1}{2}\int dp_{3}dp_{4}
\beta\gamma_{B\rho\rho}^{*}(134)
\delta (p_{1}+p_{3}+p_{4})
\beta^{-1}\bar{G}_{\rho B}(3)\beta^{-1}\bar{G}_{\rho B}(4)
(-2\bar{K}_{34})
\nonumber
\ee
\be
=\int dp_{3}dp_{4}\beta\gamma_{B\rho\rho}^{*}(134)
\delta (p_{1}+p_{3}+p_{4})\beta^{-1}\bar{G}_{\rho B}(3)
(\bar{K}_{34})
\int\frac{d\bar{\omega}_{4}}{2\pi}
\frac{\bar{\omega}_{4}\bar{G}_{\rho\rho}(\bar{\omega}_{4})}
{\omega_{4}-\bar{\omega}_{4}+i\eta}
\nonumber
\ee
\be
=\int dp_{3}dp_{4}\bar{K}_{34}
\int\frac{d\bar{\omega}_{4}}{2\pi}
\bar{\omega}_{4}\bar{G}_{\rho\rho}(\bar{\omega}_{4})
\beta\gamma_{B\rho\rho}^{*}(134)
\delta (q_{1}+k_{3}+k_{4})
\delta (\omega_{1}+\omega_{3}+\omega_{4})
\nonumber
\ee
\be
\times\beta^{-1}\bar{G}_{\rho B}(3)
\frac{1}
{\omega_{4}-\bar{\omega}_{4}+i\eta}
\nonumber
\ee
\be
=\int dk_{3}dk_{4}\bar{K}_{34}
\delta (q_{1}+k_{3}+k_{4})
\int\frac{d\bar{\omega}_{4}}{2\pi}
\bar{\omega}_{4}\bar{G}_{\rho\rho}(\bar{\omega}_{4})
\nonumber
\ee
\be
\times\int\frac{d\omega_{3}}{2\pi}
\beta\gamma_{B\rho\rho}^{*}(134)
\beta^{-1}\bar{G}_{\rho B}(3)
\frac{1}
{-\omega_{1}-\omega_{3}-\bar{\omega}_{4}+i\eta}
\nonumber
\ee
\be
=(-1)\int dk_{3}dk_{4}\bar{K}_{34}
\delta (q_{1}+k_{3}+k_{4})
\int\frac{d\bar{\omega}_{4}}{2\pi}
\bar{\omega}_{4}\bar{G}_{\rho\rho}(\bar{\omega}_{4})
\nonumber
\ee
\be
\times
\int\frac{d\omega_{3}}{2\pi}
\beta\gamma_{B\rho\rho}^{*}(134)
\beta^{-1}\bar{G}_{\rho B}(3)
\frac{1}
{\omega_{1}+\omega_{3}+\bar{\omega}_{4}-i\eta}.
\ee
Now $\bar{G}_{\rho B}(3)$ is analytic in the upper half plane.
Closing the contour in the upper half plane,
\be
\beta\Gamma_{B\rho}^{(2,C,1)}(-p_{1})
=(-1)\int dk_{3}dk_{4}\bar{K}_{34}
\delta (q_{1}+k_{3}+k_{4})
\int\frac{d\bar{\omega}_{4}}{2\pi}
\beta\bar{\omega}_{4}\bar{G}_{\rho\rho}(\bar{\omega}_{4})
\frac{(2\pi i)}{2\pi}
\nonumber
\ee
\be
\times
\beta^{-1}\bar{G}_{\rho B}(-\omega_{1}-\bar{\omega}_{4})
\beta\gamma_{B\rho\rho}^{*}(1,-1-\bar{4},\bar{4})
\nonumber
\ee
\be
=\int dk_{3}dk_{4}\bar{K}_{34}
\delta (q_{1}+k_{3}+k_{4})
(-i\bar{K}_{34})
\int\frac{d\bar{\omega}_{4}}{2\pi}
\int\frac{d\bar{\omega}_{3}}{2\pi}
\bar{\omega}_{4}\bar{G}_{\rho\rho}(\bar{\omega}_{4})
\nonumber
\ee
\be
\times
\frac{\beta\bar{\omega}_{3}\bar{G}_{\rho\rho}(\bar{\omega}_{3})}
{-\omega_{1}-\bar{\omega}_{3}-\bar{\omega}_{4})+i\eta}
\beta\gamma_{B\rho\rho}^{*}(1,-1-\bar{4},\bar{4})
\nonumber
\ee
\be
=\int dk_{3}dk_{4}\bar{K}_{34}
\delta (q_{1}+k_{3}+k_{4})
\int\frac{d\bar{\omega}_{4}}{2\pi}
\int\frac{d\bar{\omega}_{3}}{2\pi}
\beta\bar{G}_{\rho\rho}(\bar{\omega}_{4})
\bar{\omega}_{4}
\bar{\omega}_{3}
\nonumber
\ee
\be
\times\bar{G}_{\rho\rho}(\bar{\omega}_{3})
(i\bar{K}_{34})R_{-}
\beta\gamma_{B\rho\rho}^{*}(1,-1-\bar{4},\bar{4})
\nonumber
\ee
\be
=\hat{\cal O}[J_{B\rho}^{(1)}],
\ee
where we introduce the notation
\be
\hat{\cal O}[J]=
\int dk_{3}dk_{4}
\delta (q_{1}+k_{3}+k_{4})
\int\frac{d\bar{\omega}_{4}}{2\pi}
\int\frac{d\bar{\omega}_{3}}{2\pi}
\bar{G}_{\rho\rho}(\bar{\omega}_{3})
\bar{G}_{\rho\rho}(\bar{\omega}_{4})
J
\ee
and
\be
J_{B\rho}^{(1)}=
\bar{\omega}_{4}
\bar{\omega}_{3}
(i\bar{K}_{34})R_{-}
\beta\gamma_{B\rho\rho}^{*}(1,-1-\bar{4},\bar{4}).
\ee

Next we notice that
$\Gamma_{B\rho}^{(2,C,2)}(-p_{1})=\Gamma_{B\rho}^{(2,C,3)}(-p_{1})$,
so
\be
\beta\Gamma_{B\rho}^{(2+3)}(-p_{1})=-2\frac{1}{2}\int dp_{3}dp_{4}
\beta\gamma_{B\rho\rho}^{*}(134)
\delta (p_{1}+p_{3}+p_{4})
\nonumber
\ee
\be
\times \beta^{-1}\bar{G}_{\rho B}(4)\bar{G}_{\rho \rho}(3)
\beta\gamma_{\rho\rho B}(134)
\nonumber
\ee
\be
=-\int dk_{3}dk_{4}\delta (q_{1}+k_{3}+k_{4})
\int d\bar{\omega}_{3}
\beta\gamma_{B\rho\rho}^{*}(1\bar{3},-1-\bar{3})
\bar{G}_{\rho \rho}(3)
\nonumber
\ee
\be
\times
\int d\bar{\omega}_{4} \frac{\bar{\omega}_{4}\bar{G}_{\rho \rho}(\bar{4})}
{-\omega_{1}-\bar{\omega}_{3}-\bar{\omega}_{4}+i\eta}
\beta\gamma_{\rho\rho B}(1\bar{3},-1-\bar{3})
\nonumber
\ee
\be
=\hat{\cal O}[J_{B\rho}^{(2+3)}],
\nonumber
\ee
where
\be
J_{B\rho}^{(2+3)}
=\bar{\omega}_{4}R_{-}
\beta\gamma_{B\rho\rho}^{*}(1\bar{3},-1-\bar{3})
\beta\gamma_{\rho\rho B}(1\bar{3},-1-\bar{3}).
\ee
The total is thus given by
\be
J_{B\rho}=R_{-}[i\bar{\omega}_{4}\bar{\omega}_{3}\bar{K}_{34}
\beta\gamma_{B\rho\rho}^{*}(1,-1-\bar{4},\bar{4})
+\bar{\omega}_{4}
\beta\gamma_{B\rho\rho}^{*}(1\bar{3},-1-\bar{3})
\beta\gamma_{\rho\rho B}(1\bar{3},-1-\bar{3})]
\nonumber.
\ee

We next need the vertices
\be
\beta\gamma_{B\rho\rho}^{*}(1,-1-\bar{4},\bar{4})
=1+iE_{1}-i\Omega \bar{K}_{13},
\ee
\be
\beta\gamma_{B\rho\rho}^{*}(1\bar{3},-1-\bar{3})
=1+iE_{1}-i\Omega \bar{K}_{14},
\ee
and
\be
\beta\gamma_{\rho\rho B}(1\bar{3},-1-\bar{3})
=1-iE_{4},
\ee
where
\be
\Omega =\omega_{1}+\bar{\omega}_{3}+\bar{\omega}_{4}
\ee
and
\be
E_{4}=\omega_{1}\bar{K}_{14}+\bar{\omega}_{3}\bar{K}_{34}.
\ee
We then have
\be
J_{B\rho}=R_{-}[i\bar{\omega}_{4}\bar{\omega}_{3}\bar{K}_{34}
[1+iE_{1}-i\Omega \bar{K}_{13}]
+\bar{\omega}_{4}
(1+iE_{1}-i\Omega \bar{K}_{14})
(1-iE_{4})].
\ee

If we group together the terms proportional to $\Omega$, set
$R_{-}\Omega =1$, and notice that inside the integrations each term vanishes
because of odd frequency integrals over $\bar{G}_{\rho\rho}$, we then have
\be
J_{B\rho}=R_{-}[i\bar{\omega}_{4}\bar{\omega}_{3}\bar{K}_{34}
[1+iE_{1}]+\bar{\omega}_{4}(1+iE_{1})(1-iE_{4})]
\nonumber
\ee
\be
=R_{-}[1+iE_{1}][i\bar{\omega}_{4}\bar{\omega}_{3}\bar{K}_{34}
+\bar{\omega}_{4}(1-iE_{4})]
\nonumber
\ee
\be
=R_{-}[1+iE_{1}][i\bar{\omega}_{4}\bar{\omega}_{3}\bar{K}_{34}
+\bar{\omega}_{4}(1-i(\omega_{1}\bar{K}_{14}+\bar{\omega}_{3}\bar{K}_{34})]
\nonumber
\ee
\be
=R_{-}[1+iE_{1}][\bar{\omega}_{4}
(1-i\omega_{1}\bar{K}_{14}].
\ee
Inside the integrals we are free to symmetrize with respect to
$3\leftrightarrow 4$:
\be
J_{B\rho}
=R_{-}[1+iE_{1}][\frac{1}{2}[\bar{\omega}_{4}+\bar{\omega}_{3}]
-i\frac{\omega_{1}}{2}E_{1}]
\nonumber
\ee
\be
=\frac{1}{2}R_{-}[1+iE_{1}][\Omega -\omega_{1}
-i\omega_{1}E_{1}]R_{-}
\nonumber
\ee
\be
\nonumber
=\frac{1}{2}[1+iE_{1}]-\frac{\omega_{1}}{2}[1+iE_{1}]^{2}R_{-}
\ee
and
\be
\beta\Gamma_{B\rho}(-p_{1})=\hat{\cal O}[
\frac{1}{2}[1+iE_{1}]-\frac{\omega_{1}}{2}[1+iE_{1}]^{2}R_{-}].
\ee
Since $\hat{\cal O}[\frac{1}{2}iE_{1}]=0$
we have
\be
\beta\Gamma_{B\rho}^{(2,C)}(-p_{1})=\hat{\cal O}[
\frac{1}{2}-\frac{\omega_{1}}{2}[1+iE_{1}]^{2}R_{-}]
\ee
or
\be
\beta\Gamma_{B\rho}^{(2,C)}(p_{1})=\tilde{\cal O}[
\frac{1}{2}+\frac{\omega_{1}}{2}\tilde{R}_{-}[1+iE_{1}]^{2}],
\label{eq:258}
\ee
where
\be
\tilde{R}_{-}=\frac{1}{\bar{\omega}_{3}+\bar{\omega}_{4}-\omega_{1}-i\eta}
\ee
and
\be
\tilde{\cal O}[J]=
\int dk_{3}dk_{4}
\delta (-q_{1}+k_{3}+k_{4})
\int\frac{d\bar{\omega}_{4}}{2\pi}
\int\frac{d\bar{\omega}_{3}}{2\pi}
\bar{G}_{\rho\rho}(\bar{\omega}_{3})
\bar{G}_{\rho\rho}(\bar{\omega}_{4})
J.
\ee
We need the result
\be
Im ~ \beta\Gamma_{B\rho}^{(2,C)}(p_{1}) =
\frac{\omega_{1}}{2}
\hat{\cal O}[
\tilde{R}_{-}[1+iE_{1}]^{2}].
\ee

\subsection{$\Gamma_{BB}$}

We turn next to the second order self-energy
with two $B$'s. After using causality, there are five nonzero graphs.  The first is
the simplest and given by
\be
\beta^{2}\Gamma_{BB}^{(2,C,0)}(-p_{1})=-\frac{1}{2}
\int dp_{3}dp_{4}\beta\gamma^{*}_{B\rho\rho}(134)
\delta (p_{1}+p_{3}+p_{4})
\bar{G}_{\rho\rho}(3)
\bar{G}_{\rho\rho}(4)
\beta\gamma_{B\rho\rho}(134),
\nonumber
\ee
while there are four similar contributions given by
\be
\beta^{2}\Gamma_{BB}^{(2,C,1)}(-p_{1})=-\frac{1}{2}
\int dp_{3}dp_{4}\beta\gamma^{*}_{B\rho\rho}(134)
\delta (p_{1}+p_{3}+p_{4})
\beta^{-1}\bar{G}_{\rho B}(3)
\bar{G}_{\rho\rho}(4)
\beta^{2}\gamma_{BB\rho}(134),
\nonumber
\ee
\be
\beta^{2}\Gamma_{BB}^{(2,C,2)}(-p_{1})=-\frac{1}{2}
\int dp_{3}dp_{4}\beta\gamma^{*}_{B\rho\rho}(134)
\delta (p_{1}+p_{3}+p_{4})
\bar{G}_{\rho \rho}(3)
\beta^{-1}\bar{G}_{\rho B}(4)
\beta^{2}\gamma_{B\rho B}(134),
\nonumber
\ee
\be
\beta^{2}\Gamma_{BB}^{(2,C,3)}(-p_{1})=-\frac{1}{2}
\int dp_{3}dp_{4}\beta^{2}\gamma^{*}_{B\rho B}(134)
\delta (p_{1}+p_{3}+p_{4})
\bar{G}_{\rho \rho}(3)
\beta^{-1}\bar{G}_{B\rho }(4)
\beta\gamma_{B\rho \rho}(134),
\nonumber
\ee
and
\be
\beta^{2}\Gamma_{BB}^{(2,C,4)}(-p_{1})=-\frac{1}{2}
\int dp_{3}dp_{4}\beta^{2}\gamma^{*}_{BB\rho }(134)
\delta (p_{1}+p_{3}+p_{4})
\beta^{-1}\bar{G}_{B \rho}(3)
\bar{G}_{\rho\rho }(4)
\beta\gamma_{B\rho \rho}(134).
\nonumber
\ee
Then, the simplest graph can be rewritten as
\be
\beta^{2}\Gamma_{BB}^{(2,C,0)}(-p_{1})=\hat{\cal O}[J_{BB}^{(0)}]
\ee
where
\be
J_{BB}^{(0)}=-\frac{1}{2}
\delta (\Omega )
|\gamma_{B\rho\rho}(134)|^{2}
\ee
\be
=i\frac{1}{2}(R_{-}-R_{+})
[1+E_{1}^{2}].
\ee
Turning to the other four contributions, they can be grouped together
 in a way that shows their  sum is real
\be
\beta^{2}\Gamma_{BB}^{(2,C,1-4)}(-p_{1})=
\int dp_{3}dp_{4}
\delta (p_{1}+p_{3}+p_{4})
\bar{G}_{\rho\rho }(3)\bar{K}_{14}
\nonumber
\ee
\be
\times
[\gamma^{*}_{B\rho\rho }(134)
\bar{G}_{\rho B}(4)
+\gamma_{B\rho\rho }(134)
\bar{G}_{B\rho }(4)]
\nonumber.
\ee
This leads to
\be
\beta^{2}\Gamma_{BB}^{(2,C,1-4)}(-p_{1})=
\hat{\cal O}[J_{BB}^{(1-4)}]
\nonumber
\ee
where
\be
J_{BB}^{(1-4)}=2\bar{\omega}_{4}\bar{K}_{14}
\int d\omega_{4}\delta (\omega_{1}+\bar{\omega}_{3}+\omega_{4})
\left[\frac{[1+iE_{1}]}{\omega_{4}-\bar{\omega}_{4}+i\eta}
+\frac{[1-iE_{1}]}{\omega_{4}-\bar{\omega}_{4}-i\eta}\right]
\nonumber
\ee
\be
=2\bar{\omega}_{4}\bar{K}_{14}
\left[\frac{[1+i\tilde{E}_{1}]}
{-\omega_{1}-\bar{\omega}_{3}-\bar{\omega}_{4}+i\eta}
+\frac{[1-i\tilde{E}_{1}]}
{-\omega_{1}-\bar{\omega}_{3}-\bar{\omega}_{4}-i\eta}\right]
\nonumber
\ee
\be
=-2\bar{\omega}_{4}\bar{K}_{14}
\bigg\{R_{-}[1+i\tilde{E}_{1}]
+R_{+}[1-i\tilde{E}_{1}]\bigg\}
\nonumber
\ee
and
\be
\tilde{E}_{1}=E_{1}-\Omega\bar{K}_{14}.
\ee
The terms generated from $\tilde{E}_{1}$ proportional to $\Omega$
give zero contribution after integration over
$\bar{\omega}_{3}$ and $\bar{\omega}_{4}$, so
\be
J_{BB}^{(1-4)}
=-2\bar{\omega}_{4}\bar{K}_{14}
\bigg\{[R_{-}[1+iE_{1}]
+R_{+}[1-iE_{1}]\bigg\}
\nonumber
\ee
\be
=-E_{1}\bigg\{R_{-}[1+iE_{1}]
+R_{+}[1-iE_{1}]\bigg\}.
\nonumber
\ee
Combining the two contributions gives
\be
J_{BB}=
i\frac{1}{2}(R_{-}-R_{+})
[1+E_{1}^{2}]
=-E_{1}\bigg\{R_{-}[1+iE_{1}]
+R_{+}[1-iE_{1}]\bigg\}
\nonumber
\ee
\be
=\frac{i}{2}
[1+iE_{1}]^{2}R_{-}-
\frac{i}{2}
[1-iE_{1}]^{2}R_{+}
\nonumber
\ee
\be
=- Im R_{-}
[1+iE_{1}]^{2}
\nonumber.
\ee
Finally we have the result
\be
\beta^{2}\Gamma_{BB}^{(2,C)}(p_{1})=-Im \tilde{\cal O}[\tilde{R}_{-}
[1+iE_{1}]^{2}]
\ee
and we see that this set of self-energies satisfies the FDT as
\be
Im ~ \Gamma_{B\rho}^{(2,C)}(p_{1})
=-\frac{\beta\omega_{1}}{2}\Gamma_{BB}^{(2,C)}(p_{1}).
\ee

\section{Statics and an effective Potential Approach}

Let us summarize our basic results up to now.  The two-point vertex is given
in the form
\be
\Gamma_{ij}=\gamma_{ij}^{(0)}+\gamma_{ij}^{(1)}
+\Gamma_{ij}^{(1,c)}
+\gamma_{ij}^{(2,s)}
+\Gamma_{ij}^{(2,c)}
\ee
where $\gamma_{ij}^{(0)}$ is the noninteracting  gas result,
$\gamma_{ij}^{(1)}$ and $\Gamma_{ij}^{(1,C)}$ are
the first order single-particle and collective contributions (respectively) to
the self-energy, and $\gamma_{ij}^{(2,s)}$ and  $\Gamma_{ij}^{(2,C)}$
are the single-particle and collective contributions at second order
in perturbation theory.
Most of the focus in this paper is on the collective part of the
two-point vertex at second order.  We will be able to show that
the collective part determines the static structure to this order
and determines the ENE phase diagram. The single-particle
contribution does not enter into either determination.
However, we anticipate that it plays a crucial role in the slow kinetics
near the ENE transition. It will be treated in detail in a
companion paper.

Let us now determine the static structure in the second order
 approximation.
This is very conveniently done in the case where there is a
linear fluctuation-dissipation theorem.  The key result is
that the zero frequency limit of the response function gives
the static structure factor
\be
G_{\rho B}(q, 0)=-\int\frac{d\omega}{2\pi}\beta G_{\rho\rho}(q,\omega )=
-\beta S(q)
\ee
in the ergodic phase.  Then, using the Dyson's equation we have
\be
G_{\rho B}(q, 0)=\frac{1}{-1/(\beta \rho_{0})-V(q)+\Gamma_{B\rho}(q,0)}
\ee
or
\be
S(q)=\frac{\rho_{0}}{1+\rho_{0}\beta V(q)-\beta\rho_{0} \Gamma_{B\rho}(q,0)}.
\ee
The needed self-energy is given by Eq.(\ref{eq:258}) as
\be
\beta\Gamma_{B\rho}(q,0)=\beta\tilde{\cal O}\bigg[\frac{1}{2}\bigg]
\nonumber
\ee
\be
=\frac{1}{2\rho_{0}^{4}}\int d3 d4 (2\pi )^{d}\delta (-q_{1}+k_{3}+k_{4})
\bar{G}_{\rho\rho}(3)
\bar{G}_{\rho\rho}(4)
\nonumber
\ee
\be
=\frac{1}{2\rho_{0}^{4}}\int \frac{d^{d}k_{3}}{(2\pi )^{d}}
 \frac{d^{d}k_{4}}{(2\pi )^{d}}
(2\pi )^{d} \delta (q_{1}+k_{3}+k_{4})
\int \frac{d\omega_{3}}{2\pi}\bar{G}_{\rho\rho}(3)
\int \frac{d\omega_{4}}{2\pi}
\bar{G}_{\rho\rho}(4).
\ee
We showed previously that
\be
\int \frac{d\omega_{3}}{2\pi}\bar{G}_{\rho\rho}(3)=
\bar{G}_{\rho\rho}(k_{3},t=0)=-\rho_{0}\beta V(k_{3})S(k_{3}),
\ee
therefore,
\be
\beta\Gamma_{B\rho}(q,0)=
\frac{1}{2\rho_{0}^{4}}\int \frac{d^{d}k_{3}}{(2\pi )^{d}}
 \frac{d^{d}k_{4}}{(2\pi )^{d}}
(2\pi )^{d} \delta (q_{1}+k_{3}+k_{4})
\rho_{0}\beta V(k_{3})S(k_{3})\rho_{0}\beta V(k_{4})S(k_{4}).
\nonumber
\ee
Then, after introducing dimensionless wavenumber $k_{3}'=k_{3}\sigma$
(where $\sigma$ is
the hard-sphere diameter), $\tilde{S}(q')=S(q)/\rho_{0}$,
$\tilde{V}(q')=\rho_{0}\beta V(q)$, and
\be
\tilde{S}(q')=\frac{1}{1+\tilde{V}(q')-M(q')},
\label{eq:278}
\ee
where
\be
M(q')=\rho_{0}\beta \Gamma_{B\rho}(q,0)
=\frac{\pi}{12\eta}\int \frac{d^{d}k_{3}'}{(2\pi )^{d}}
 \frac{d^{d}k_{4}'}{(2\pi )^{d}}
 \delta (q_{1}'+k_{3}'+k_{4}')
\tilde{V}(k_{3}')\tilde{S}(k_{3}')\tilde{V}(k_{4}')\tilde{S}(k_{4}'),
\label{eq:278a}
\ee
where
$\eta =\pi \rho \sigma^{3}/6$
is the packing fraction for hard spheres.  For simplicity we drop
the primes on the dimensionless wavenumbers.

So far the analysis has been rather general.  Let us apply
these results to the case of three-dimensional hard spheres.
We immediately have a problem if we view our perturbation theory
expansion in the conventional way since the Fourier transform
of a hard-core potential is not well defined. We can however
take a different approach.  Instead of taking $V(q)$ as given and computing
$S(q)$, we take $S(q)$ as given and, at a given order, determine the effective
potential by inverting the equivalent of Eq.(\ref{eq:278}).
Thus, at first order
\be
\tilde{S}(q)=\frac{1}{1+\tilde{V}(q)}
 \ee
and the effective potential is essentially the direct correlation function given by
\be
\tilde{V}(q)=\frac{1}{\tilde{S}(q)}-1.
\ee

We assume that the structure factor for a hard-sphere system
is given by the solution\cite{PY} to the Percus-Yevick equation.
In Fig. 1 we plot the structure factor for different packing fractions.

\begin{figure}[btp]
\centering
\subfigure[]{\includegraphics[width=.5\textwidth]{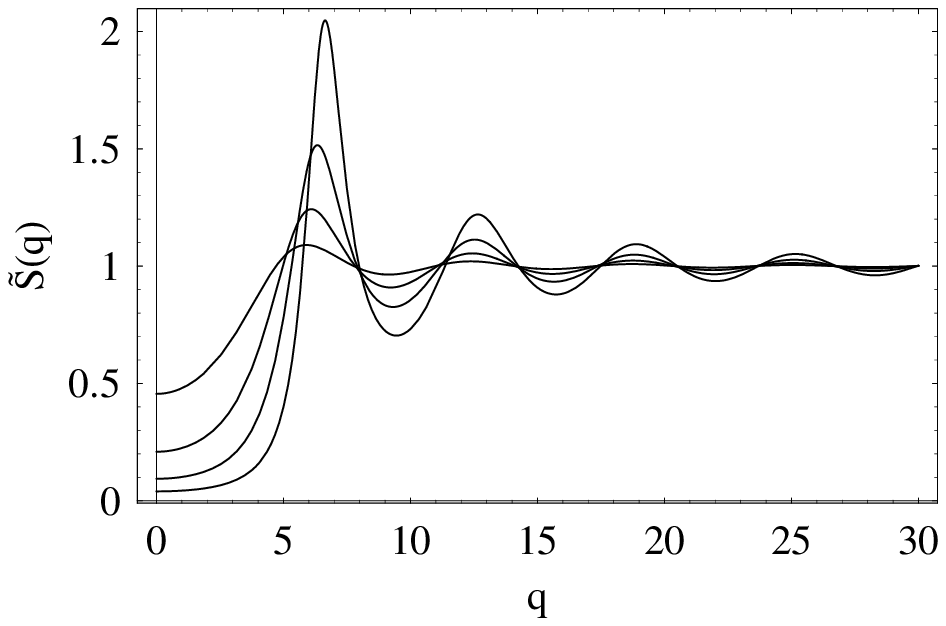}}
\subfigure[]{\includegraphics[width=.5\textwidth]{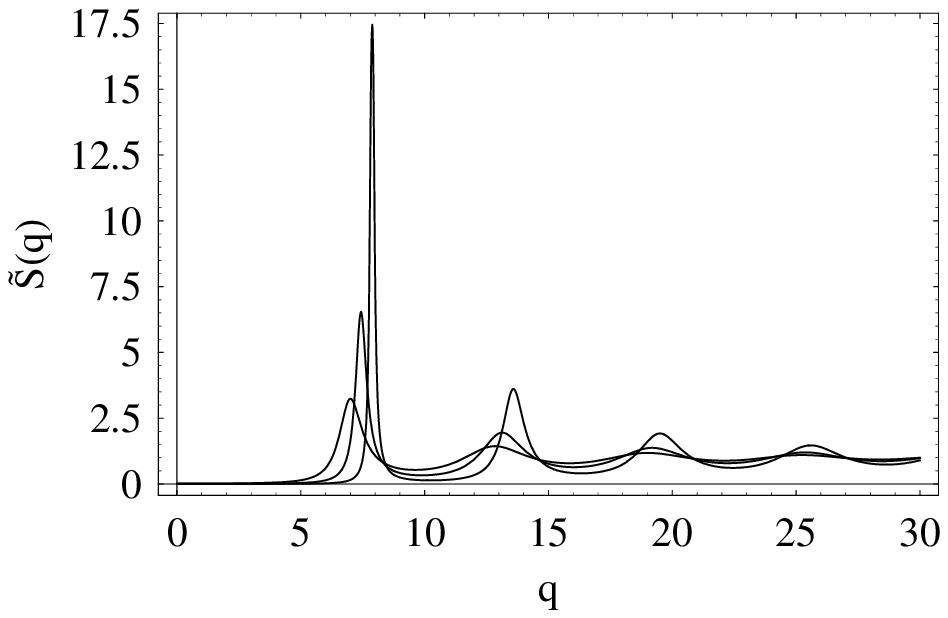}}
\subfigure[]{\includegraphics[width=.5\textwidth]{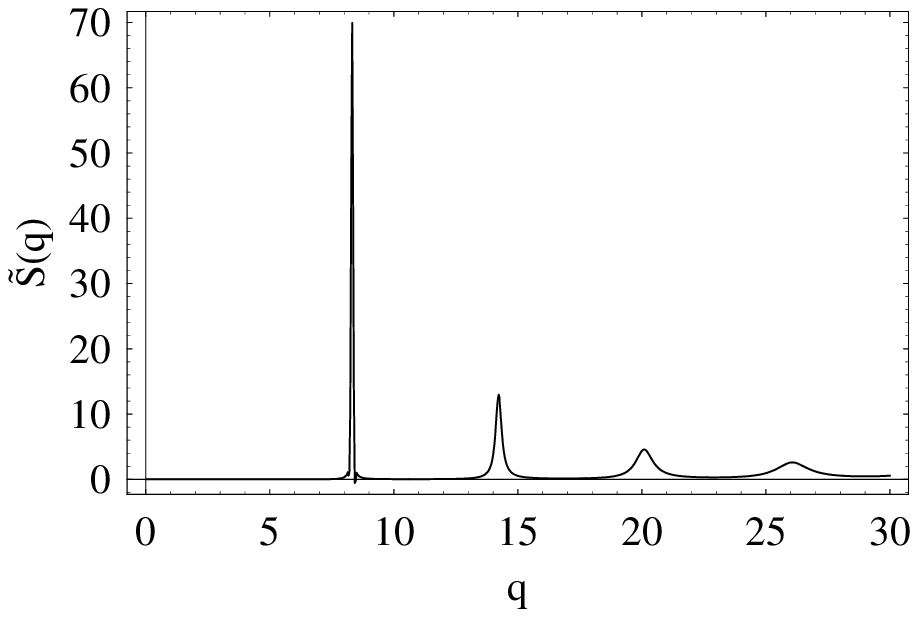}}
\caption{The static structure factor as determined using the Percus-Yevick Hard Sphere Approximation for (a) $\eta = 0.1$, $0.2$, $0.3$, and $0.4$, (b) $\eta = 0.5$, $0.6$ and $0.7$, and (c) $\eta=0.8$. Larger first structure peak height corresponds to larger $\eta$ within each plot.}
\end{figure}

The numerical solution of Eq.(\ref{eq:278})
for $\tilde{V}(q)$ is obtained as follows. First, rewrite Eq.(\ref{eq:278}) in the form
\be
\tilde{V}=\tilde{S}^{-1}-1+M(\tilde{V}).
\ee
Next, define
\be
V_{0}=\tilde{S}^{-1}-1
\ee
and rewrite
\be
\tilde{V}=V_{0}+Q(\tilde{V}),
\ee
where
\be
Q(\tilde{V})=M[V_{0}+Q(\tilde{V})]\equiv I(Q).
\ee
We seek an iterated solution. To this end, we write
\be
\alpha Q_{\ell +1}+(1-\alpha )Q_{\ell}=I(Q_{\ell})
\ee
with the number $\alpha$ chosen appropriately. This
equation can be rewritten as a functional recursion relation,
\be
Q_{\ell +1}= Q_{\ell}+\frac{1}{\alpha}[I(Q_{\ell})-Q_{\ell}].
\ee
At a fixed point we have a solution to the original problem.
By trial and error we find that a reasonable choice for $\alpha$
is $\alpha = 100.$
Then beginning with $Q_{0}=0$ we monitor
\be
\Delta_{\ell}=4\frac{\sum_{q}(\tilde{V}_{\ell +1}-\tilde{V}_{\ell})^{2}}
{\sum_{q}(\tilde{V}_{\ell +1}+\tilde{V}_{\ell})^{2}}.
\ee
Over the entire range of $\eta$ studied, we find that $\Delta_{\ell}$
is driven to small values ($10^{-6}$-$10^{-8}$) with a minimum
controlled by the range of wavenumber included in the calculation.
One obtains good accuracy if one chooses $q_{max}=80$.

In Fig.2 we give the results for the effective potential for a
sequence of densities.  Also shown is $V_{0}$.  A key quantity
is $\tilde{V}\tilde{S}$.  It is plotted in Fig. 3 where we also plot
$V_{0}\tilde{S}$.
Clearly the two quantities are close and do not change dramatically
in going from first to second order. The biggest shift is for small
wavenumbers. That $M(q)$ has a maximum for small $q$ is understood
by assuming that $\tilde{V}(k)\tilde{S}(k)$ is sharply peaked for
$k = k_{0}$.  Therefore
$\tilde{V}(k)\tilde{S}(k)\tilde{V}(k+q)\tilde{S}(k+q)$ is small except
when $ k\approx |q+k|\approx k_{0}$ which is true for small $q$.

\begin{figure}[btp]
\centering
\subfigure[]{\includegraphics[width=.45\textwidth]{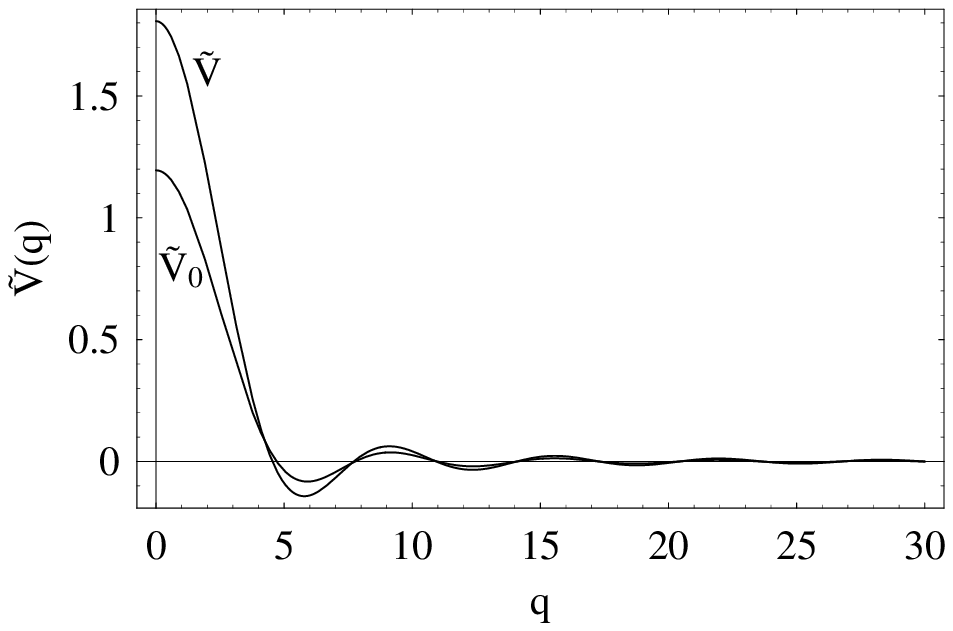}}
\subfigure[]{\includegraphics[width=.45\textwidth]{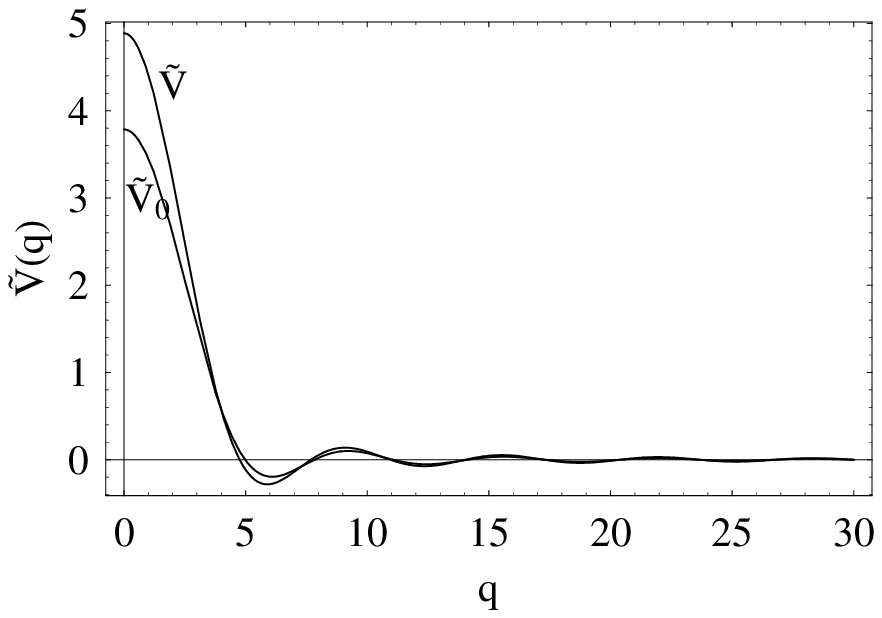}}
\subfigure[]{\includegraphics[width=.45\textwidth]{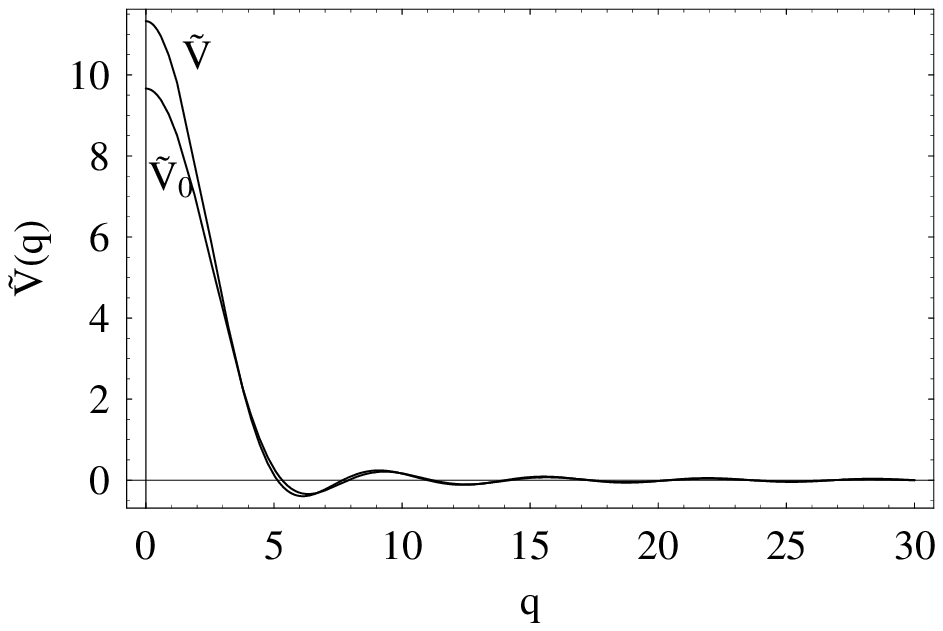}}
\subfigure[]{\includegraphics[width=.45\textwidth]{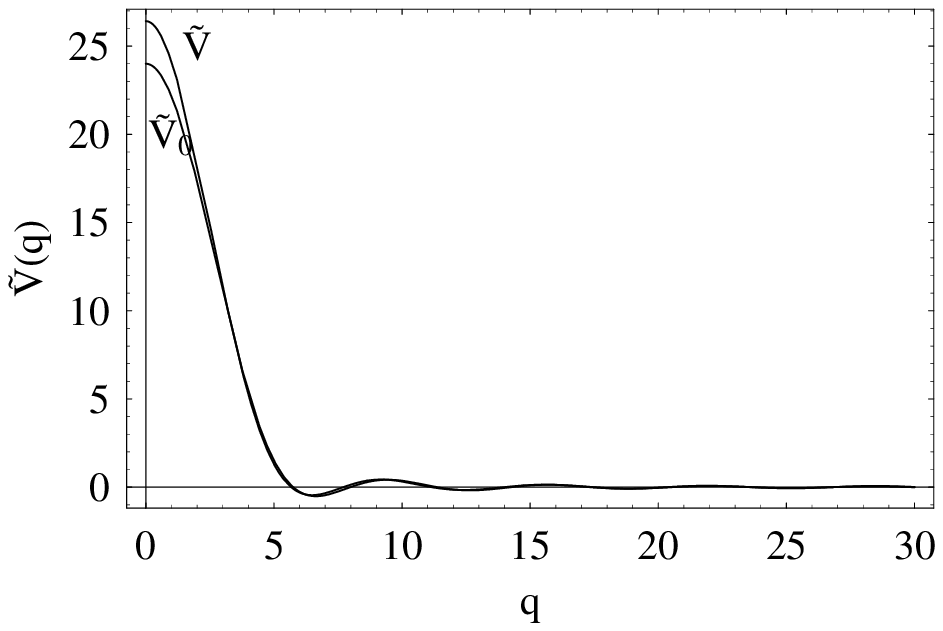}}
\subfigure[]{\includegraphics[width=.45\textwidth]{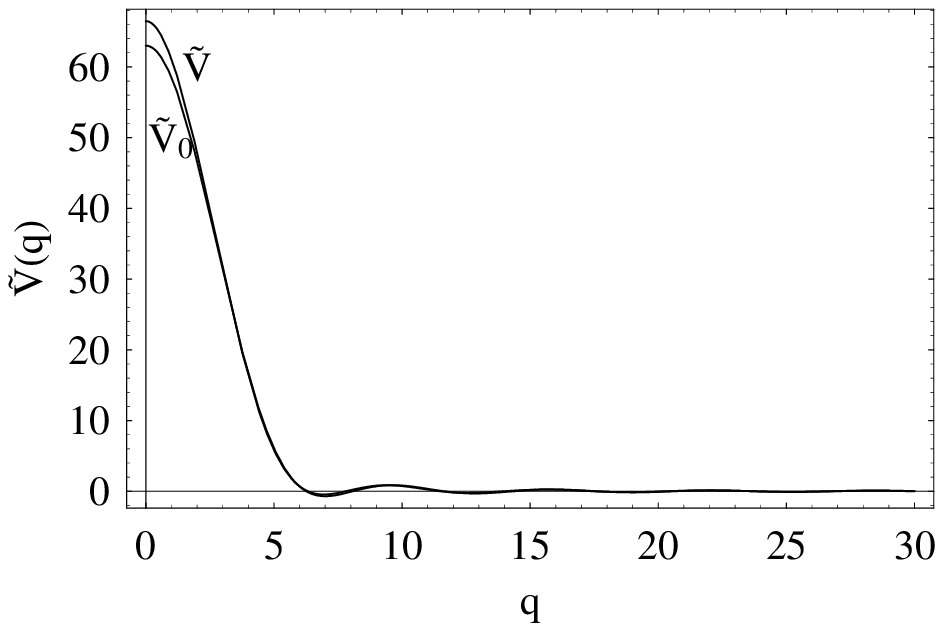}}
\subfigure[]{\includegraphics[width=.45\textwidth]{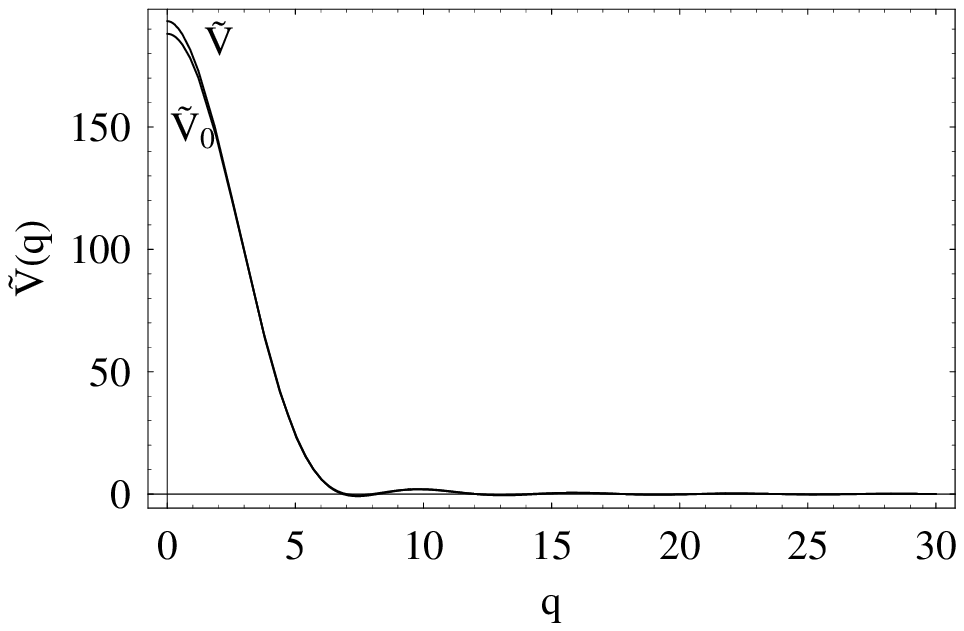}}
\subfigure[]{\includegraphics[width=.45\textwidth]{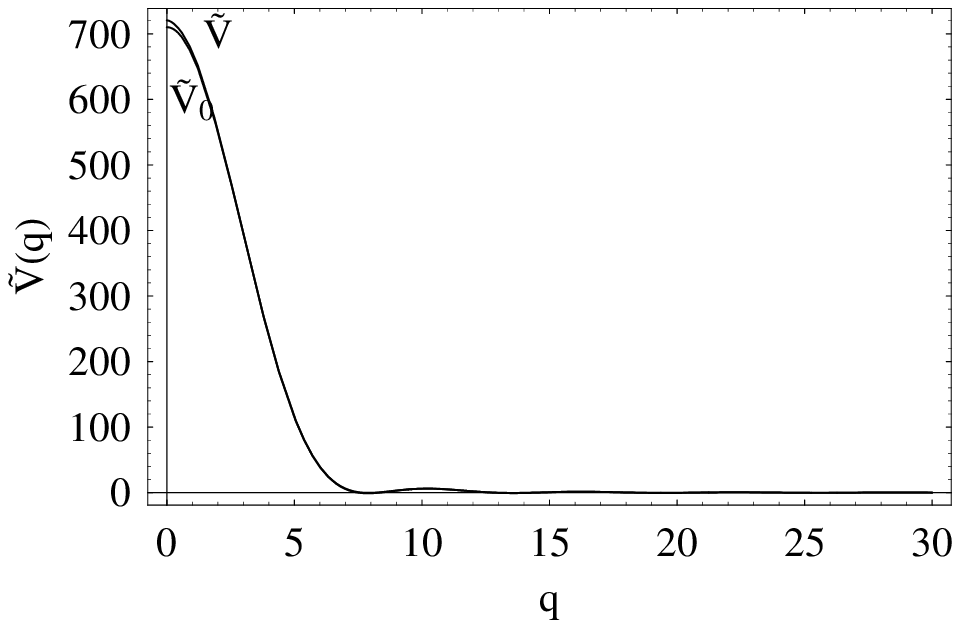}}
\subfigure[]{\includegraphics[width=.45\textwidth]{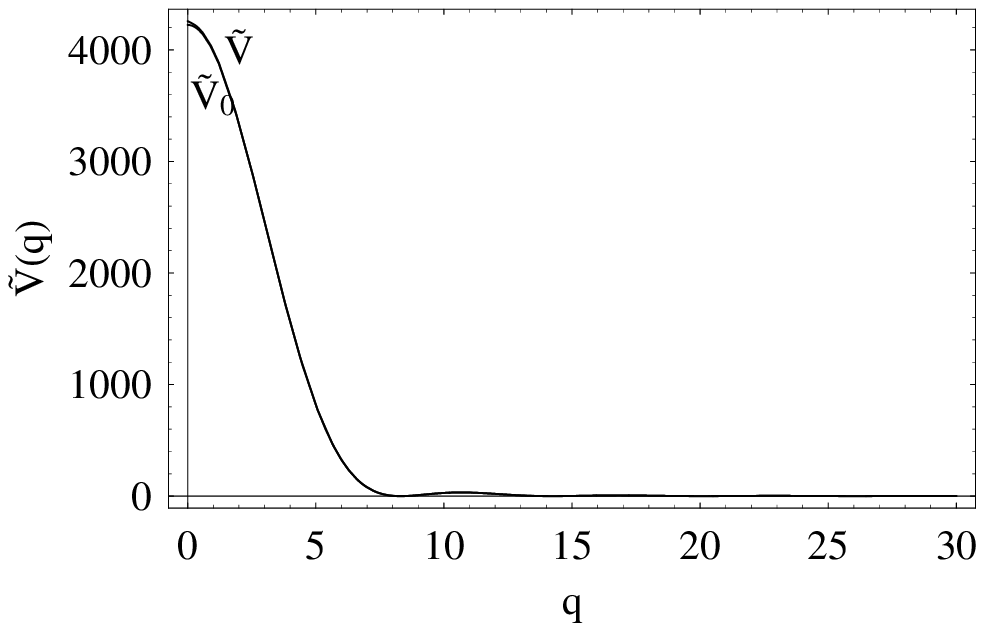}}
\caption{The zeroth order potential $\tilde{V}_0(q)$ and the numerical solution for the full potential $\tilde{V}(q)$ from (a) $\eta = 0.1$ to (h) $\eta=0.8$ in increments of 0.1.}
\end{figure}

\begin{figure}[btp]
\centering
\subfigure[]{\includegraphics[width=.45\textwidth]{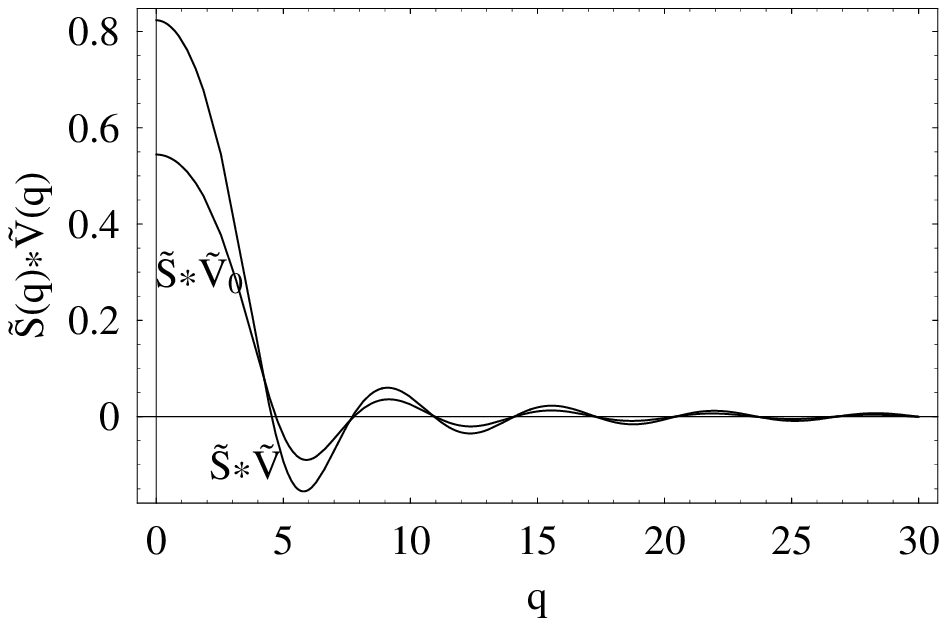}}
\subfigure[]{\includegraphics[width=.45\textwidth]{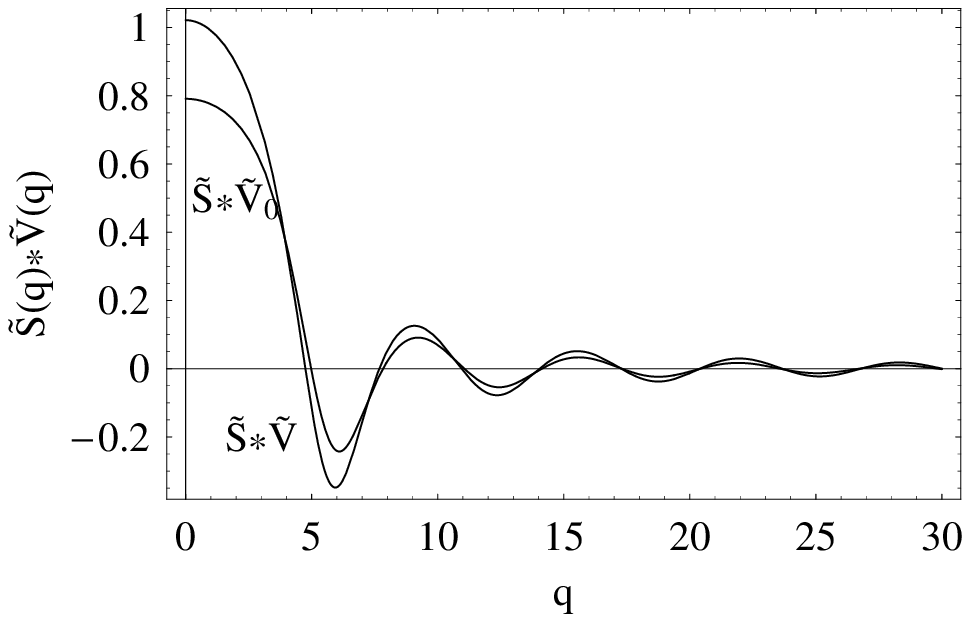}}
\subfigure[]{\includegraphics[width=.45\textwidth]{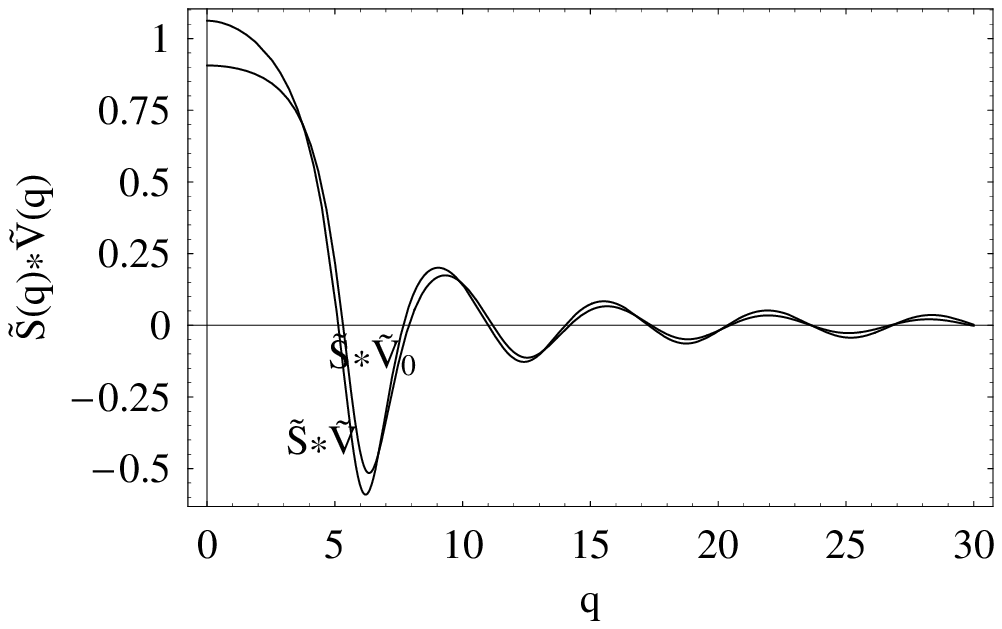}}
\subfigure[]{\includegraphics[width=.45\textwidth]{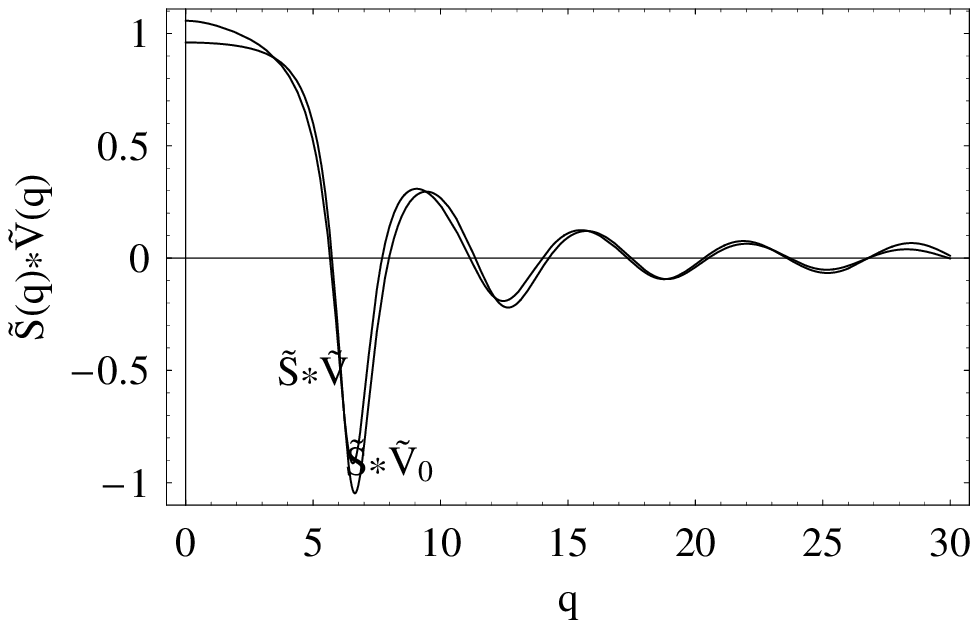}}
\subfigure[]{\includegraphics[width=.45\textwidth]{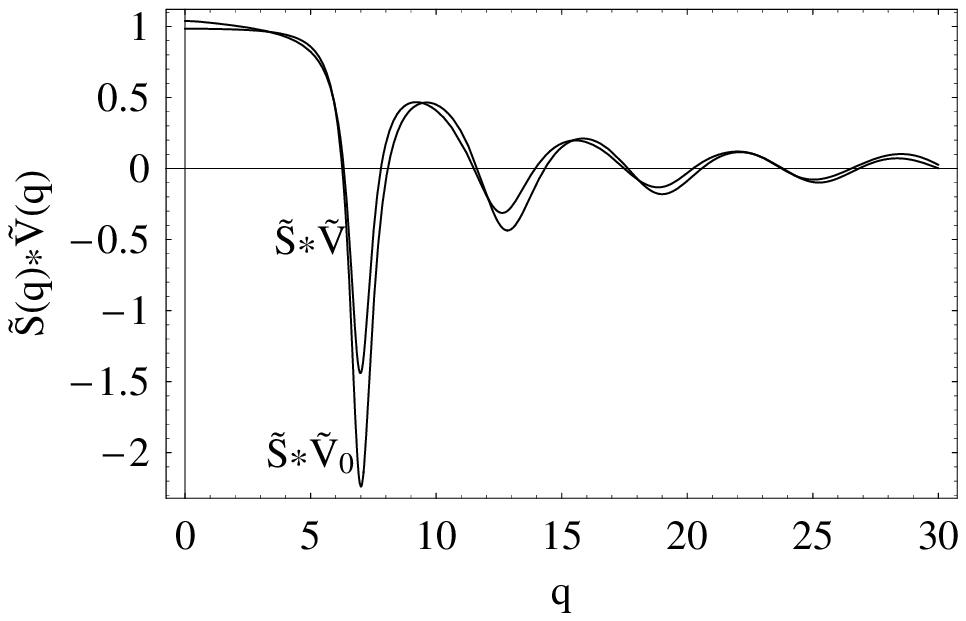}}
\subfigure[]{\includegraphics[width=.45\textwidth]{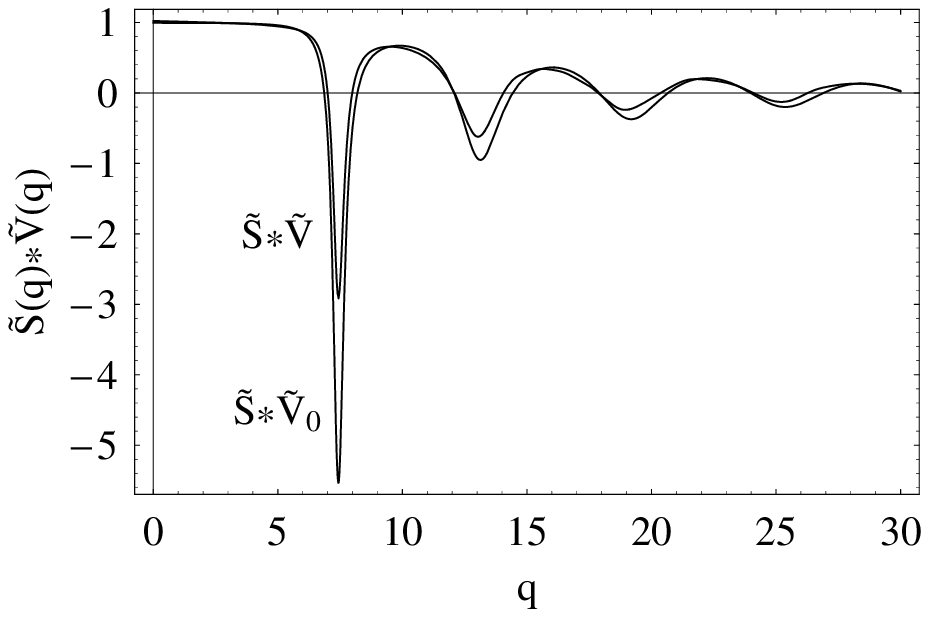}}
\subfigure[]{\includegraphics[width=.45\textwidth]{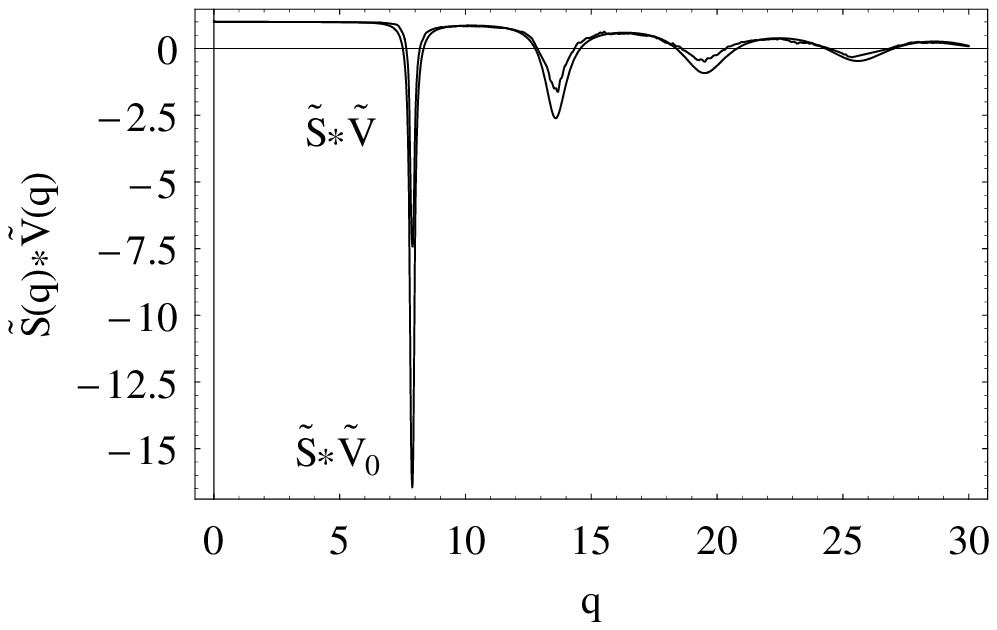}}
\subfigure[]{\includegraphics[width=.45\textwidth]{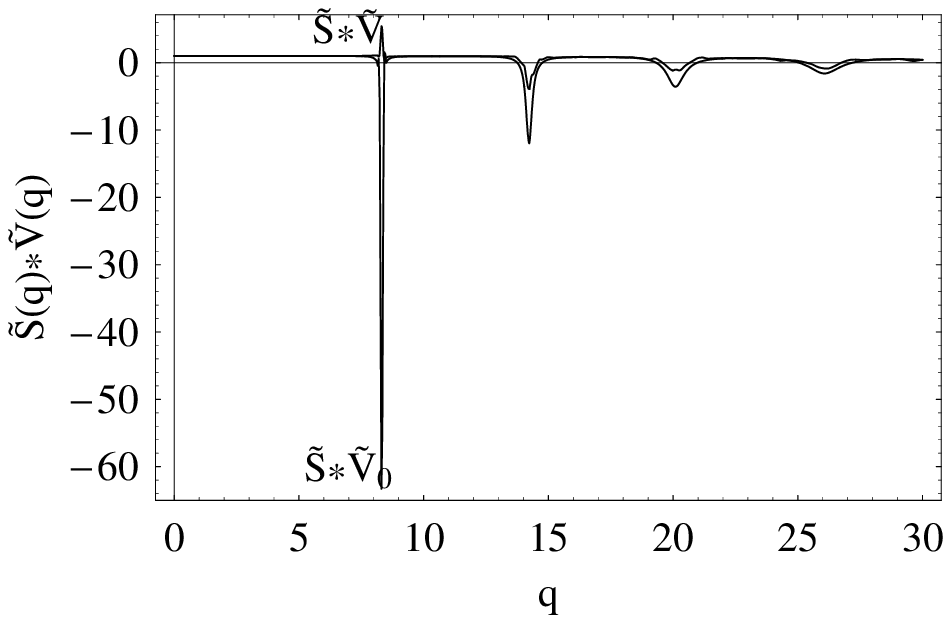}}
\caption{The potential times the static structure factor for both $\tilde{V}_0(q)$ and $\tilde{V}(q)$ from (a) $\eta = 0.1$ to (h) $\eta=0.8$ in increments of 0.1.}
\end{figure}

\section{ENE Transition}

As we increase the density, (or lower the temperature), our model
system slows down.  It is possible that there is a density
above which the density-density correlation does not decay
to zero for long times.
In the nonergodic phase
\be
\lim_{t\rightarrow\infty}G_{\rho\rho}(q,t)=F(q) > 0
\nonumber.
\ee
In terms of the time Fourier transform,
\be
G_{\rho\rho}(q,\omega )=F(q)2\pi \delta (\omega )+
g_{\rho\rho}(q,\omega ),
\label{eq:297}
\ee
where $g_{\rho\rho}$ is the {\it regular} contribution.
In the nonergodic phase, assuming the FDT still holds,
one has for
the response function
\be
G_{\rho B}(q,\omega =0)=\int_{-\infty}^{\infty}G_{\rho B}(q,t)
\nonumber
\ee
\be
=\beta\int_{-\infty}^{\infty}\theta (t)
\frac{\partial}{\partial t}G_{\rho\rho }(q,t)
\nonumber
\ee
\be
=\beta\int_{0}^{\infty}
\frac{\partial}{\partial t}G_{\rho\rho }(q,t)
=\beta [F(q)-S(q)]
\label{eq:509}
\ee
while
\be
G_{B\rho }(q,\omega =0)=\int_{-\infty}^{\infty}G_{B\rho }(q,t)
\nonumber
\ee
\be
=\beta\int_{-\infty}^{\infty}\theta (-t)(-1)
\frac{\partial}{\partial t}G_{\rho\rho }(q,t)
\nonumber
\ee
\be
=\beta\int_{-\infty }^{0}(-1)
\frac{\partial}{\partial t}G_{\rho\rho }(q,t)
=-\beta S(q)
\label{eq:512}.
\ee

There are several equivalent ways of determining the nonergodicity
parameter $F$ as a function of density and temperature. One approach
developed in the time-domain results from pursuing the memory-function
equation, Eq.(\ref{eq:101}).  This will be explored elsewhere.
Here we give a more direct
complementary analysis in frequency space.

Using one of the Dyson equations we have
\be
G_{\rho\rho}(q,\omega )=-G_{\rho B}(q, \omega )\Gamma_{BB}(q,\omega )
G_{B\rho}(q ,\omega ).
\label{eq:299}
\ee
For an ENE transition we  associate a zero-frequency delta-function
in $\Gamma_{BB}$
\be
\Gamma_{BB}( q,\omega )=-2\pi\delta (\omega )\bar{\Gamma}(q)
+\textrm{regular part}.
\label{eq:300}
\ee
Then, using Eqs.(\ref{eq:297}) and (\ref{eq:300}) in Eq. (\ref{eq:299}),
we find
\be
F(q)=\beta^{2}\bar{\Gamma}(q)[S(q)-F(q)]S(q)
\label{eq:301}.
\ee
Equation (\ref{eq:301}) can be rewritten as
\be
\frac{F(q)}{S(q)-F(q)}=S(q)\beta^{2}\bar{\Gamma}(q).
\nonumber
\ee
We need to extract the lowest order approximation for
$\bar{\Gamma}$.  We have from our previous work that
\be
\beta^{2}\Gamma_{BB}^{(2)}(1)= Im {\cal O}[R_{-}(1+iE_{1})^{2}]
\nonumber
\ee
\be
=-\rho_{0}^{-4}\int\frac{d\bar{\omega}_{3}}{2\pi}
\frac{d\bar{\omega}_{4}}{2\pi}
\frac{d^{d}\bar{k}_{3}}{(2\pi )^{d}}
\frac{d^{d}\bar{k}_{4}}{(2\pi )^{d}}
\delta (q+\bar{k}_{3}+\bar{k}_{4})
\bar{G}_{\rho\rho}(\bar{k}_{3},\bar{\omega}_{3})
\bar{G}_{\rho\rho}(\bar{k}_{4},\bar{\omega}_{4})
\nonumber
\ee
\be
\times
Im {\cal O}[R_{-}(1+iE_{1})^{2}].
\nonumber
\ee
In the nonergodic phase we have the $\delta$-function contribution
\be
\bar{G}_{\rho\rho}(\bar{k}_{3},\bar{\omega}_{3})
=-\rho_{0}\beta V(\bar{k}_{3})F(\bar{k}_{3})2\pi\delta (\bar{\omega}_{3})
\ee
and we find immediately that we can set $E_{1}=0$, and
\be
Im R_{-}=\pi\delta (\omega_{1}+\omega_{3}+\omega_{4})
=\pi \delta (\omega_{1}).
\ee

Comparing with Eq.(\ref{eq:300}),
\be
\beta^{2}\bar{\Gamma}_{B}(q)=
\frac{1}{2\rho_{0}^{4}}\int \frac{d^{d}k}{(2\pi )^{d}}
\rho_{0}\beta V(k) F(k)
\rho_{0}\beta V(q+k) F(q+k)
\nonumber.
\ee
It is useful to write
\be
F(q)=S(q)f(q)
\ee
and it is $f(q)$ that is conventionally called the
ergodicity  parameter.

Using the same set of dimensionless variables
as in the static calculations, the nonergodicity equation
can be written in the conventional form
\be
\frac{f(q)}{1-f(q)}=w(q),
\label{eq:305}
\ee
where
\be
w(q)=\frac{\pi}{12\eta}\tilde{S}(q)
\int \frac{d^{d}k}{(2\pi )^{d}}
\tilde{V}(k) f(k)\tilde{S}(k)
\tilde{V}(q+k)f(k+q)\tilde{S}(q+k).
\label{eq:306}
\ee

Before solving Eq.(\ref{eq:305}) numerically, it is useful to obtain an
approximate analytical solution.  Notice that the combination
$\tilde{V}(k) \tilde{S}(k)$ (see Fig. 3) is sharply
peaked at the structure factor maximum.  Then, to a reasonable
approximation, we can replace $f(k)$ with its value at the maximum such that
\be
w(q)=f^{2}(q_{0})\tilde{S}(q)M(q)
\ee
with $M(q)$ the same quantity that appears in the static calculation.
Putting this result in the nonergodicity equation we obtain
\be
\frac{f(q)}{1-f(q)}=f^{2}(q_{0})\tilde{w}(q),
\label{eq:308}
\ee
where
\be
\tilde{w}(q)=\tilde{S}(q)\tilde{M}(q)
\ee
is known from our static structural calculations.  We now have a closed
algebraic equation for $f(q_{0})$ if we set $q=q_{0}$
in Eq.(\ref{eq:308}) with $f_{0}=f(q_{0})$
and $w_{0}=\tilde{w}(q_{0})$:
\be
\frac{f_{0}}{1-f_{0}}=f_{0}^{2}w_{0}.
\ee
Discarding the ergodic solution $f_{0}=0$, we have a quadratic
equation to solve given by
\be
w_{0}f_{0}^{2}-w_{0}f_{0}+1=0
\ee
with the solution
\be
f_{0}=\frac{1}{2}\left(1+\sqrt{1-4/w_{0}}\right).
\ee
Note that there is no transition for $w_{0}<4$.  For $w_{0}> 4$, we obtain
the full wave number dependence by putting $f_{0}$ back into
\be
f(q)=\frac{\tilde{w}f_{0}^{2}}{1+\tilde{w}f_{0}^{2}}.
\label{eq:295}
\ee

It is easy to compute the $w_{0}$ numerically using the Percus Yevick (PY) structure
factor.  In Fig. 4 we plot $w_{0}$ versus $\eta$.  We find that the
ENE transition density, when $w_{0}=4$, is $\eta^{*}=0.53$.  The
associated $f(q)$ is shown in Fig. 5 for several densities in the
nonergodic phase.

\begin{figure}[btp]
\centering
\subfigure[]{\includegraphics[width=.48\textwidth]{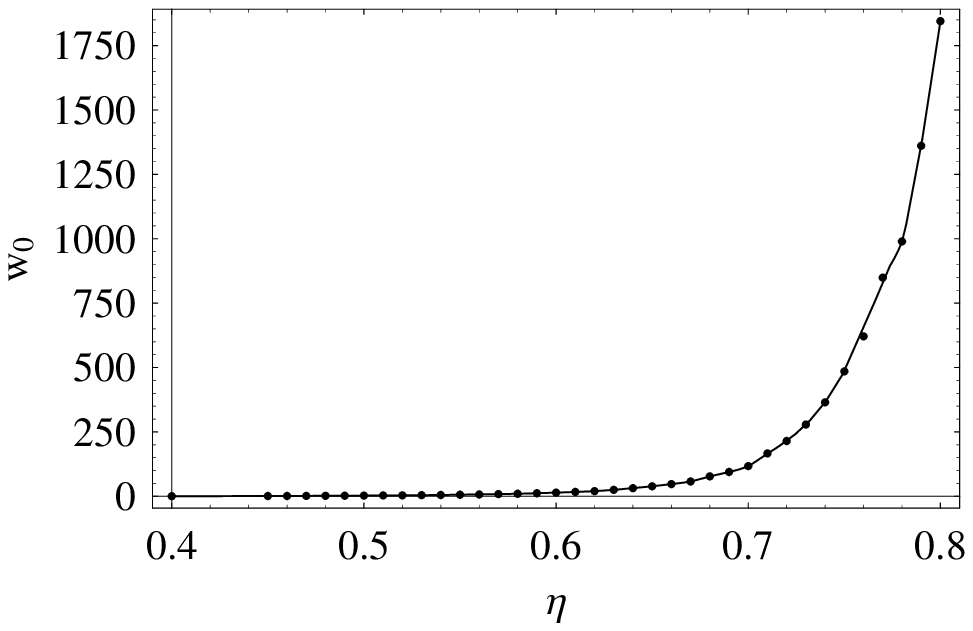}}
\subfigure[]{\includegraphics[width=.48\textwidth]{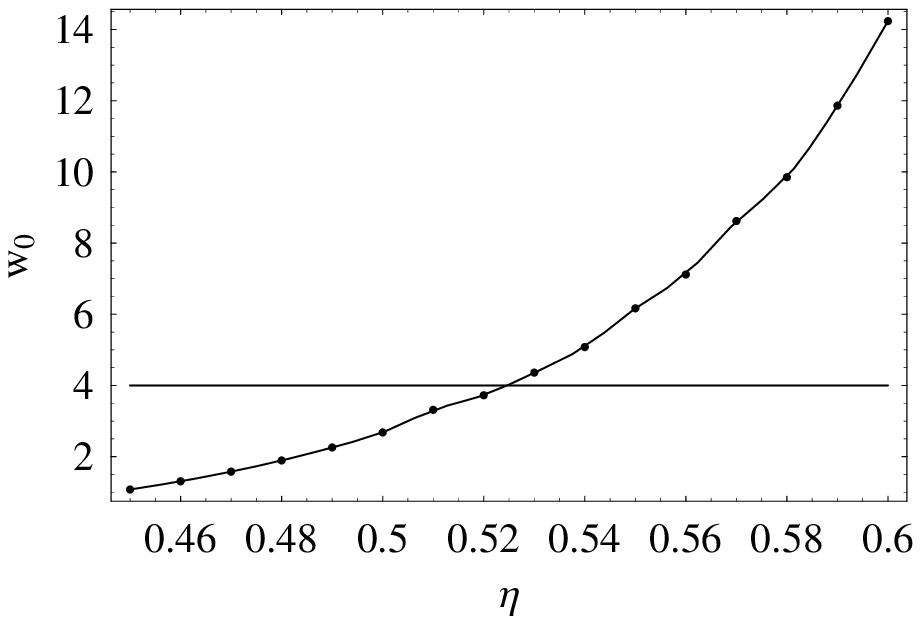}}
\caption{Values of $w_0$ versus $\eta$. Note that the line crosses $w_0=4$ (and therefore enters the nonergodic regime) near $\eta = 0.525$. (The solid line is to guide the eye only.)}
\end{figure}

\begin{figure}[btp]
\centering
\subfigure[]{\includegraphics[width=.45\textwidth]{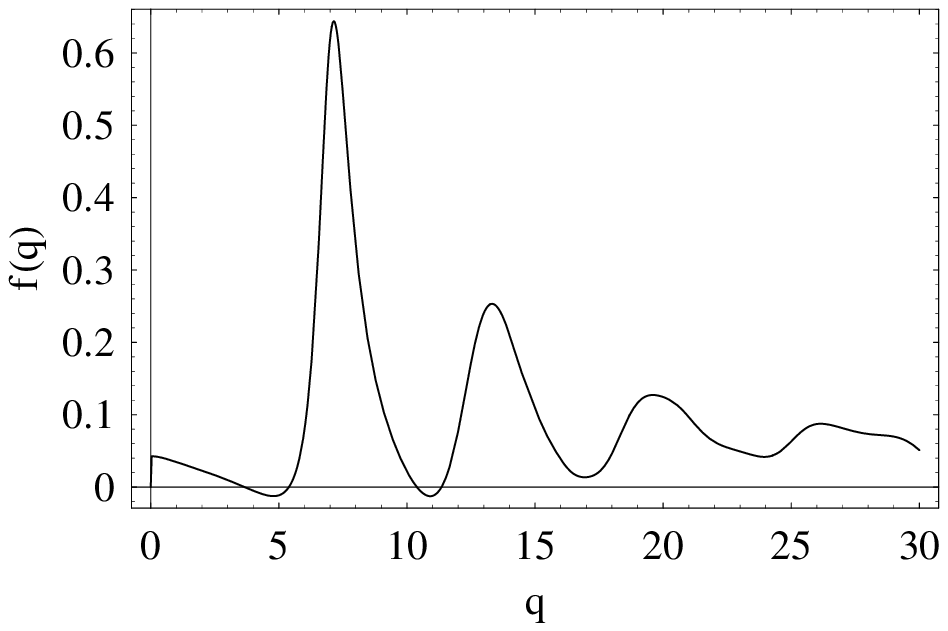}}
\subfigure[]{\includegraphics[width=.45\textwidth]{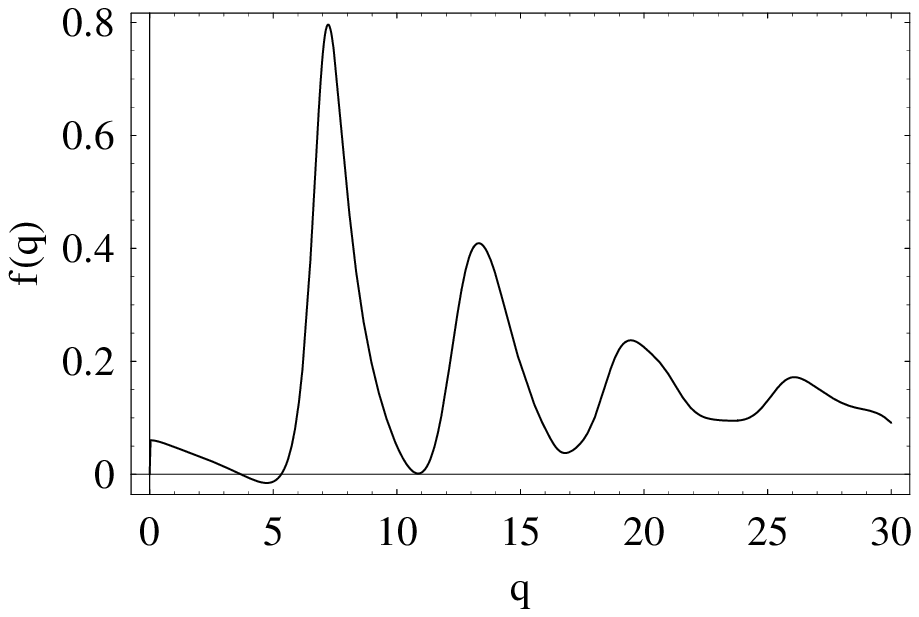}}
\subfigure[]{\includegraphics[width=.45\textwidth]{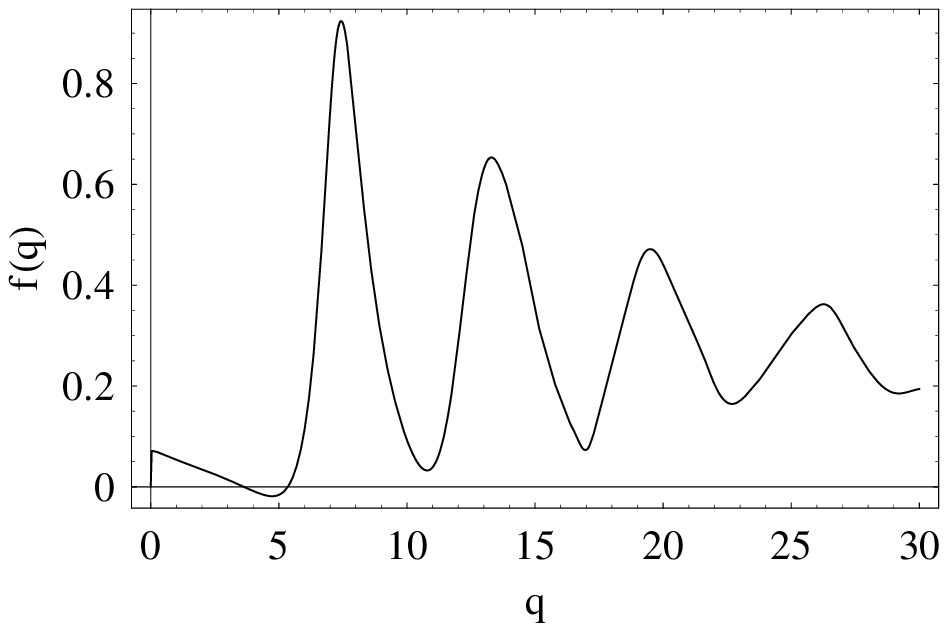}}
\subfigure[]{\includegraphics[width=.45\textwidth]{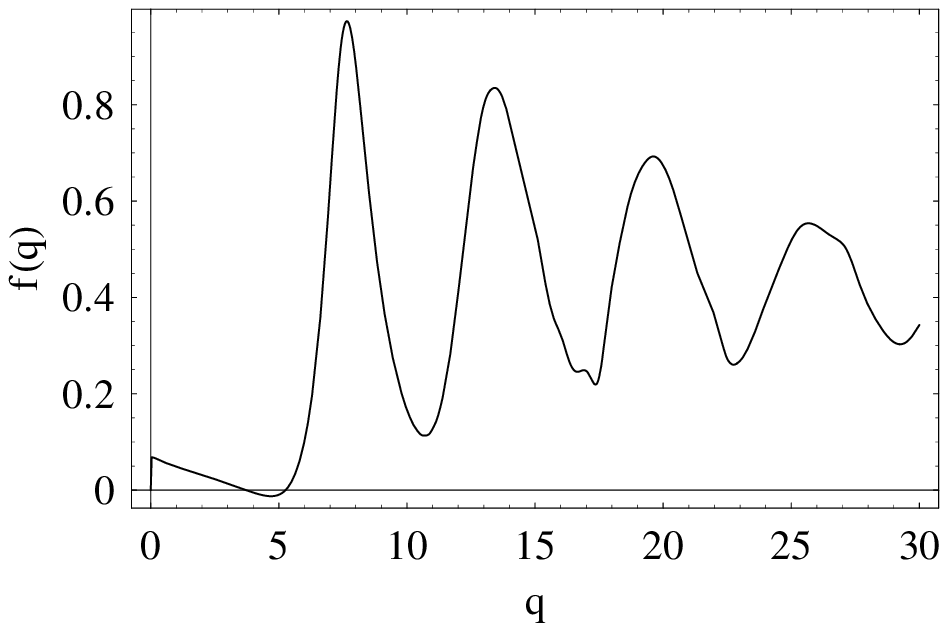}}
\subfigure[]{\includegraphics[width=.45\textwidth]{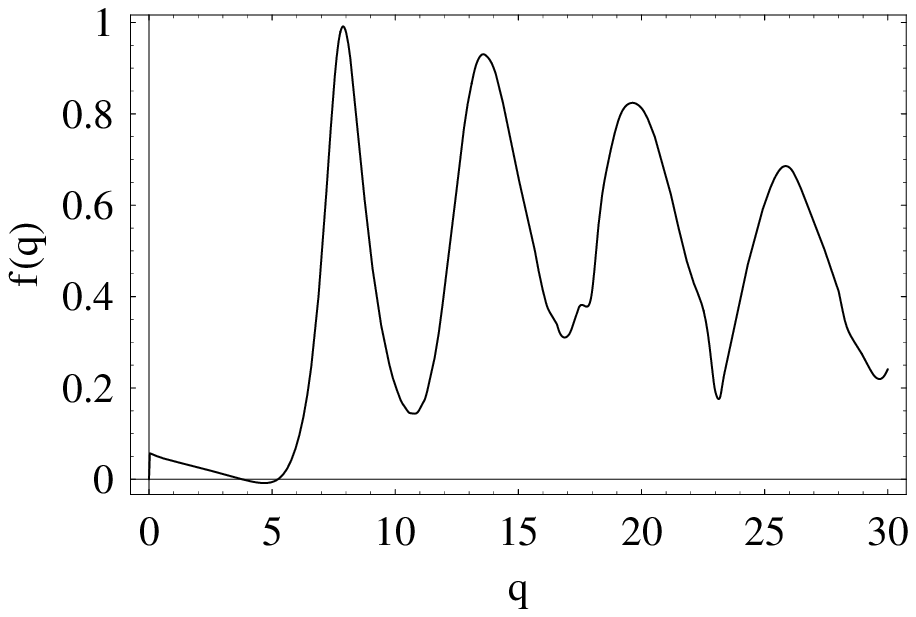}}
\subfigure[]{\includegraphics[width=.45\textwidth]{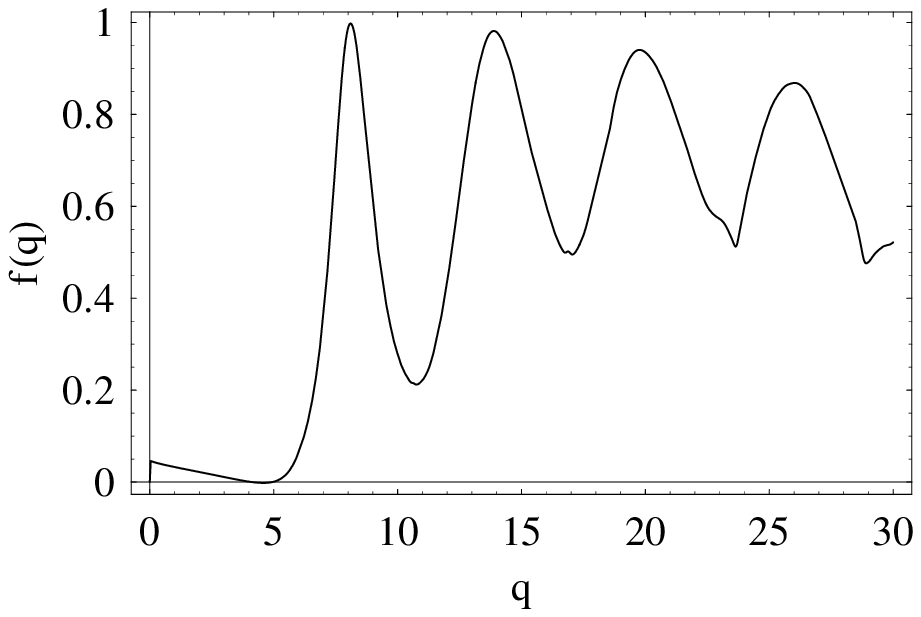}}
\caption{Using the analytic approximation for f(q) given by Eq. (\ref{eq:295}), we plot the approximate form of the nonergodicity parameter for (a) $\eta=0.53$ (just above the analytic transition), (b) $\eta=0.55$, (c) $\eta=0.60$, (d) $\eta=0.65$, (e) $\eta=0.70$ and (f) $\eta=0.75$.}
\end{figure}

Returning to the full problem,
Eq.(\ref{eq:305}) can be rewritten in the form
\be
f(q)=\frac{w(q)}{1+w(q)}.
\label{eq:340}
\ee
The solution of this equation is available via direct iteration.
If we use an initial trial value of $f_0(q)=1/2$
in Eq.(\ref{eq:306}), we generate an initial trial value $w_0=0.25\tilde{S}(q)M(q)$
to obtain
\be
f_{1}(q)=w_{0}/(1+w_{0}).
\ee
We continue iterating via
\be
f_{\ell +1}(q)=w_{\ell}/(1+w_{\ell})
\ee
and find that
\be
\lim_{\ell\rightarrow \infty}(f_{\ell +1}-f_{\ell})=0.
\ee
For pure hard-sphere systems in three dimensions we find an
ergodic-nonergodic transition at $\eta^{*}=0.76$.
Clearly this density is physically unattainable.
It is interesting to solve Eq.(\ref{eq:340}) using the first
order result $\tilde{V}=\tilde{S}^{-1}-1$ rather than the second
order result.  Despite the fact that there is small change in
the effective potential in going from first to second order
one finds a substantial change in the critical density from
$0.76$ to $0.60$.  Thus, we see that $\eta^{*}$ is a sensitive
quantity.

In Fig. 6
we plot $f(q)$ for hard spheres for a set of $\eta > \eta^{*}$.
We find a lot of structure in $f(q)$.
Comparing the approximate $f(q)$ given in Fig. 5 with the full
numerical solution we see good agreement despite very different transition
densities.

\begin{figure}[btp]
\centering
\subfigure[]{\includegraphics[width=.48\textwidth]{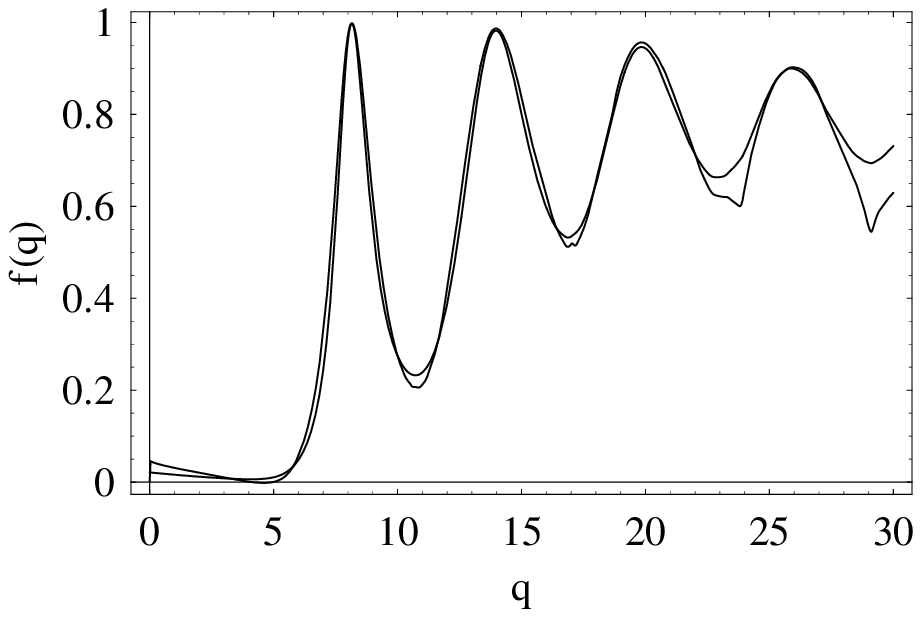}}
\subfigure[]{\includegraphics[width=.48\textwidth]{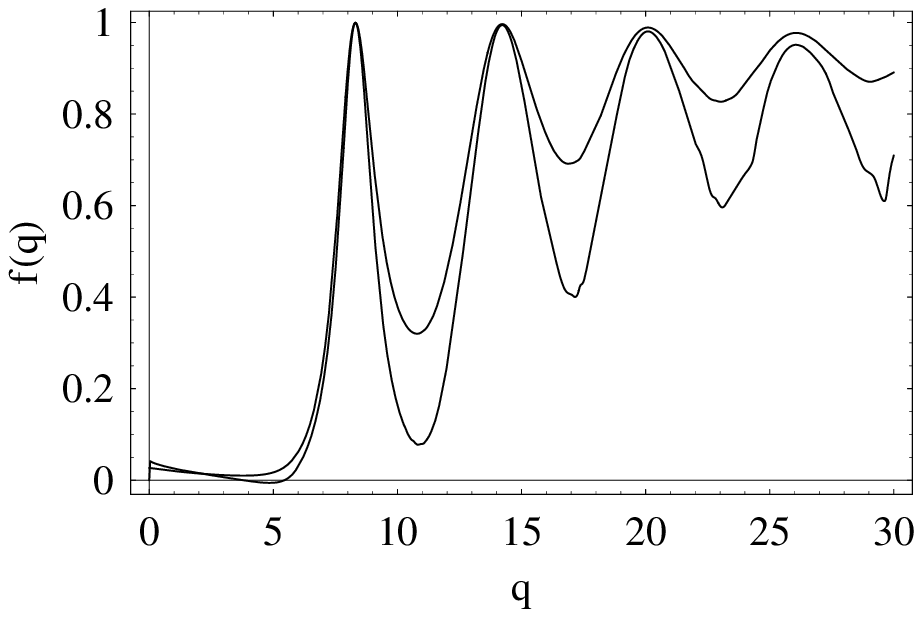}}
\caption{Numerically, we find a transition at $\eta=0.763$. We plot the numerical solution and the analytic solution for (a) $\eta=0.763$ and (b) $\eta=0.8$. The more ragged and slightly lower line in each plot is the analytic approximation.}
\end{figure}

\section{Conclusions}

We established here that the theory of
time-dependent fluctuations  in equilibrium
can be organized around the fluctuation-dissipation
theorem.
Using the FDT one can relate the two-point response function
to the two-point correlation function.
In turn this gives a linear-relation between self-energy components.
We show more specifically the nontrivial result that the collective
parts of the self-energies at second order satisfy a FDT.  We find
the second order self-energies as functionals of the physical
density-density correlation functions.  The role of the three-point
vertex functions are crucial in this analysis.

Using the FDT the static structure is separated from the dynamical
information and gives one a self-consistent expression for the
static structure factor in terms of the potential. This
expression can be recovered from a strictly static analysis.
We show how the problem can be turned around and posed as in MCT.
The structure factor is assumed to be given and we solve for the
effective potential $\tilde{V}$ that produces the known structure factor.
At first order, the effective potential is proportional to the
static direct correlation function. At second order, $\tilde{V}$ is
similar to the first order result except at small wavenumbers.

We also show that the theory is compatible with an ENE transition
at second order.  The critical density is sensitive to the details of
the calculation.  Using the Percus-Yevick approximation for
hard spheres we find an ENE transition
at the physically inaccessible density $\eta^{*}=0.76$.

In a future work we will explore the role of the single-particle
dynamics which occur in this theory. The treatment of these degrees of
freedom gives one information about the equation of state,
$\langle\rho\rangle =\bar{\rho}(\rho_{0},T)$.
The single-particle degrees of freedom contribute a second order
contribution to the self-energies which depends  on the full
density-density correlation function.

Our focus here, because of its simplicity, has been on
Smoluchowski dynamics, but as will be discussed elsewhere,
the method developed here can be applied to Newtonian
dynamics and Fokker-Planck dynamics as well as a broader
class of models.

The approach presented here allows for a systematic method for
analyzing corrections to this
second order result including determination of
higher order correlation functions.
We plan to analyze the third order contribution to the collective
self-energy soon.

We guess that the theory presented here can be organized to give
a theory of freezing.  One can then do simultaneous free
energy comparisons between the nonergodic state and the crystalline
solid state.

\begin{acknowledgements}
I thank the Physics Department and the Joint Theory Institute at the University
of Chicago for support. A special thanks is due to M. Zannetti for showing that $G_{\rho B}$
satisfies a simple linear fluctuation-dissipation theorem and for other comments
on this work. I thank David McCowan for help with the figures and comments and I thank
Paul Sprydis and S. Das for their comments on this work.
\end{acknowledgements}


\begin{thebibliography}{99}
\bibitem{SD}  The history of Smoluchowski dynamics is discussed
in FTSPD.
\bibitem{Brownian}
A brief discussion of the storied history of Brownian motion
is given in Refs. 2 and 20 in FTSPD. We will refer to the
noninteracting limit of Smoluchowski dynamics as the
Brownian gas.
\bibitem{MSR1}
MSR actions are discussed in Appendix A in FTSPD.
\bibitem{OZR}
The static Ornstein-Zernike
relation
that connects the
radial distribution function to the direct correlation function.
\bibitem{TDDC} In different contexts the kinetic kernels have
different names.  In the field theory context the kernels are
typically called self-energies, in the kinetic theory case,
where the analysis is in terms of retarded quantities, the kernel is
called a memory function, and in the general case it can be called
a dynamic direct correlation function.
\bibitem{multim}  Note that $b_{1}$
depends on the $n$ $t_{i}$ while the operator depends only on $t_{1}$.
\bibitem{FTP} An advantage of working in equilibrium is that we can
take the initial time to minus infinity.  We called this the
field-theory protocol in FTSPD.
\end{thebibliography}
\end{document}